\documentclass[%
 reprint,
 amsmath,amssymb,
 aps,
prb,
floatfix,
]{revtex4-2}
\usepackage{mathtools}
\usepackage{bm}
\usepackage{graphicx}% Include figure files
\usepackage{dcolumn}% Align table columns on decimal point
\usepackage{bm}% bold math
\usepackage{epstopdf}
\usepackage{xcolor}
\usepackage{xspace}

\usepackage{hyperref}% add hypertext capabilities
\usepackage[all]{hypcap}
\hypersetup{colorlinks, linkcolor=blue,          % color of internal links (change box color with linkbordercolor)
    citecolor=blue,        % color of links to bibliography
    filecolor=blue,      % color of file links
    urlcolor=blue           % color of external links
}

\newcommand{\mub}{\mu_\text{B}}

\newcommand{\mo}{\mu_0}
\newcommand{\Tn}{T_\text{N}}
\newcommand{\CRO}{Cs$_2$RuO$_4$\xspace}
\newcommand{\CCB}{Cs$_2$CoBr$_4$\xspace}

\begin{document}

\preprint{APS/123-QED}

\title{Spin-flop-like transition as quantum critical point in \CRO}% Force line breaks with \\

\author{S.D. Nabi$^1$}
 \email{nabid@ethz.ch}
\author{M. Zhu$^1$}%
\author{K. Yu. Povarov$^2$}%
\author{D. G. Mazzone$^3$}%
\author{J. Lass$^3$}%
\author{Y. Wu$^4$}
\author{Z. Yan$^1$}%
\author{S. Gvasaliya$^1$}%
\author{A. Zheludev$^1$}%
 \email{http://www.neutron.ethz.ch/}
\affiliation{%
 $^1$Laboratory for Solid State Physics, ETH Zurich, 8093 Z\"urich, Switzerland\\
 $^2$Dresden High Magnetic Field Laboratory (HLD-EMFL) and W\"urzburg-Dresden Cluster of Excellence ct.qmat, Helmholtz-Zentrum Dresden-Rossendorf (HZDR), 01328 Dresden, Germany\\
 $^3$PSI Center for Neutron and Muon Sciences, Paul Scherrer Institute, 5232 Villigen, Switzerland\\
 $^4$Neutron Scattering Division, Oak Ridge National Laboratory, Oak Ridge, Tennessee 37831, USA}%

\date{\today}% It is always \today, today,
             %  but any date may be explicitly specified

\begin{abstract}
We report thermodynamic, neutron diffraction, and inelastic neutron scattering measurements on \CRO, a member of the celebrated family of frustrated magnets Cs$_2MX_4$ ($M$ = Cu, Co, $X$ = Br, Cl). Unlike the previously studied members, it is based on 4$d$ transition metal ions with $S=1$. Mapping out the $H-T$ magnetic phase diagram reveals an unusual continuous spin-flop-like phase transition associated with a quantum critical point within the antiferromagnetically ordered phase. A quantitative analysis of the complex magnetic excitation spectrum measured in zero field allows us to derive a model magnetic Hamiltonian for this compound. Its main feature is a frustration of magnetic anisotropy on a level that is much higher than in any of the previously studied species. This frustration naturally explains the peculiar phase transition observed.

\end{abstract}

%\keywords{Suggested keywords}%Use showkeys class option if keyword
                              %display desired
\maketitle

%\tableofcontents

\section{\label{sec:level1}Introduction\protect\\}
Geometrically frustrated quantum antiferromagnets offer the best chance of discovering exotic quantum magnetic phases and excitations. Materials with the general formula Cs$_2MX_4$ ($M$ = transition metal, $X$ = halogen) are among the oldest known families of frustrated magnets. Perhaps the most extensively studied species is the $S=1/2$-based Cs$_2$CuCl$_4$. This compound realizes the frustrated $J$-$J'$ distorted triangular lattice model. It was shown to bridge the exotic  physics of one-dimensional (1D) spin chains with that of a two-dimensional (2D) triangular lattice \cite{CCC1, CCC2, CCC3, CCC4}. The isostructural Cs$_2$CuBr$_4$ is famous for a cascade of exotic magnetic phase transitions and plateau states induced by external magnetic fields \cite{CCB1, CCB2}. Members of the family that contain $S=3/2$ Co$^{2+}$ ions, such as Cs$_2$CoCl$_4$ \cite{CCoC1, CCoC2, CCoC3}, feature frustration due to a misalignment of local magnetic anisotropy axes in addition to frustration of magnetic interactions. The more recently studied \CCB \cite{Povarov2020, Facheris2022, Facheris2023, Facheris2024} is best described as a frustrated zig-zag spin ladder system \cite{Facheris2024}. It is host to one of the most spectacular known examples of Zeeman ladder hierarchies of spinon bound states \cite{Facheris2023}.

In the present work we introduce an entirely new member of this structural family, namely \CRO. In place of 3$d$ ions it features   Ru$^{6+}$. In this 4$d$ transition metal, we can expect an enhancement of relativistic spin-orbit coupling effects. Moreover, it is a $S=1$ ion, a first for this group of materials. Additionally, having O$^{2-}$ ions in place of halogen $X^-$ can be expected to drastically affect the superexchange pathways and thereby give rise to a completely novel hierarchy of magnetic interactions. In the present work we explore the magnetic properties of this exciting material and construct a minimal model spin Hamiltonian that can account for them. Our most interesting finding is a peculiar field-induced quantum critical point (QCP) located {\em inside} the magnetically ordered phase. We attribute it to an unusual {\em continuous} spin flop transition enabled by competing single-ion anisotropies.

\section{\label{sec:level1}Methods and Material\protect\\}

\begin{figure}[t]
\includegraphics[clip, trim=0.0cm 0.0cm 0.0cm 0.0cm ,width=9 cm]{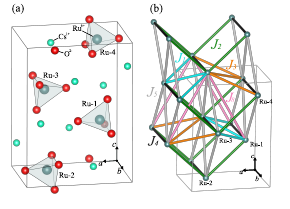}% Here is how to import EPS art
\caption{\label{fig:crystal_structure} Crystal structure and magnetic exchange paths in \CRO. (a) Schematic overview of the unit cell, indicated by a black line. The four magnetic ions per unit cell are labeled by Ru-1 to Ru-4. (b) Schematic overview of the relevant symmetry allowed magnetic exchange paths. The exchange pathways are labeled $J_1$-$J_5$ and $J_8$.}
\end{figure}

Like  Cs$_2MX_4$ compounds, \CRO crystallizes in an orthorhombic structure with space group $Pnma$ (No. 62). The lattice parameters are $a = 8.528(2)$, $b = 6.494(2)$, and $c = 11.498(2)$~\AA. Dark green crystals were grown by the self-flux method in a platinum crucible. The crystal structure was validated using single crystal x-ray diffraction on a Bruker APEX-II diffractometer. The structural refinement with anisotropic thermal factors (quality factor $R=2.4$~\%) was based on the analysis of 1116 independent Bragg reflections. The results are summarized in Table~\ref{tab:tablecrys}. They are consistent with those of previous studies \cite{Fischer1990}. A schematic illustration of the atomic arrangement is given in Fig.~\ref{fig:crystal_structure}(a). The magnetism stems from the four equivalent Ru$^{6+}$ ions ($S$ = 1) in each unit cell. Their local coordination environment is a distorted tetrahedron formed by four O$^{2-}$ anions. The relevant superexchange pathways between magnetic ions are discussed in detail in Ref.~\cite{Facheris2024}.
The corresponding exchange constants $J_1$--$J_5$ span distances between 5 and 7~\AA, and are shown in Fig.~\ref{fig:crystal_structure}(b). The most prominent structural feature is chains of magnetic ions that run along the $b$ axis, bonded by $J_4$. The next shell of interactions $J_6$--$J_8$ involves  larger distances, between 8.5 and 8.7~\AA. For reasons clarified below, the bond $J_8$ is also shown in the figure.

\begin{table}[t]
\caption{\label{tab:tablecrys}%
Crystal structural parameters for \CRO determined at room temperature using single crystal x-ray diffraction.
}
\begin{ruledtabular}
\begin{tabular}{l|c|c|c|c}
\textrm{Atom} & \textrm{$x$} & \textrm{$y$} & \textrm{$z$} & \textrm{$U_{\text{eq}}$} \\
\colrule
Cs 1 & 0.98962(5) & 0.25     & 0.20442(3) & 0.02384(15) \\
Cs 2 & 0.83324(5) & 0.75     & 0.41277(3) & 0.02697(16) \\
Ru 1 & 0.27361(6) & 0.75     & 0.42061(3) & 0.01721(16) \\
O 1  & 0.2003(7)  & 0.75     & 0.5654(4)  & 0.0317(10)  \\
O 2  & 0.4801(7)  & 0.75     & 0.4184(6)  & 0.0476(15)  \\
O 3  & 0.1954(5)  & 0.5270(5) & 0.3524(3)  & 0.0368(8)  \\
\end{tabular}
\end{ruledtabular}
\end{table}

Single crystal samples of \CRO were characterized in a series of magnetic and thermodynamic experiments. Magnetic susceptibility was measured on a 0.25~mg sample using a Quantum Design Magnetic Property Measurement System (MPMS) SQUID Magnetometer. These were acquired in a temperature range from 1.8~K to 300~K, with a probing field of $\mo H$ = 0.1~T along the principal crystallographic directions. Magnetization was measured at 1.8~K between 0 and 7~T for the same field orientations. Low-temperature magnetization was measured on the same sample using a Faraday-balance capacitive magnetometer \citep{fb} at 250~mK and 2~K, with magnetic field up to 9~T along the $b$ axis. Using the same setup, magnetic torque was measured in fields up to 9~T and temperatures ranging from 200~mK to 4~K. The torque signal corresponds to the deflection of a miniature cantilever on which the sample is mounted. The magnetic field sweep rate was optimized to minimize eddy current heating. Additionally, high field magnetization curves along all crystallographic directions were collected using a compensated pickup coil setup in pulsed fields up to 35~T at the High Magnetic Field Laboratory in Dresden (HLD-EMFL) \cite{HMFL}. Both low-temperature and high-field magnetization data were calibrated to absolute units using those collected on the MPMS. Heat capacity was measured on the same sample using a Quantum Design Physical Property Measurement System (PPMS) in conjunction with a $^3$He--$^4$He dilution refrigerator (DR) insert. Data were collected using the standard relaxation method. The experimental temperature range spanned 100~mK to 8~K, and magnetic fields up to 12.8~T. The field was applied along two orthogonal crystallographic directions $a$ and $b$. For the measurements in the DR range (between 100~mK and 2~K), the crystal was mounted on a silver-foil sample holder to fix the orientation.

Nuclear neutron diffraction measurements have been performed at the two-axis neutron diffractometer ORION at the Paul Scherrer Institute (PSI) with an incoming neutron wavelength of $\lambda$ = 3.3~\AA~at room temperature.
Magnetic neutron diffraction experiments were conducted at the thermal diffractometer DEMAND \cite{demand} at the High Flux Isotope Reactor (HFIR) at Oak Ridge National Laboratory (ORNL). A 40~mg sample of \CRO was installed in the $(0, k, l)$ scattering plane and mounted inside a standard orange cryostat, reaching a base temperature of 1.5~K. Neutron wavelength $\lambda$ = 1.533~\AA~was selected using a vertical focusing Si(220) monochromator. To suppress higher-order contamination, a pyrolytic graphite (PG) filter was placed in the incident beam path. The diffraction data was processed with Python scripts at the MantidWorkbench platform. Neutron spectroscopy experiments were performed on a 1~g sample of \CRO at the multiplexing spectrometer CAMEA \citep{CAMEARSI} at PSI. The sample was installed with the $(0, k, l)$ scattering plane horizontal and mounted in an orange cryostat. Foreground measurements were collected at a base temperature of $T$ = 1.6~K. Six measurement series were performed using incoming neutron energies of $E_i$ = 5~meV, 5.13~meV, 5.04~meV, 5.08~meV (elastic resolution $\sim$~0.19~meV ), 4.38~meV, and 4.45~meV (elastic resolution $\sim$~0.14~meV) — the latter two providing higher energy resolution at low energy transfers. Sequential steps of approximately 0.08~meV were used to interlace the datasets and suppress horizontal intensity artifacts. Each incoming energy was measured at 2$\theta$ = -41$^{\circ}$, -45$^{\circ}$. For each 2$\theta$, the sample was rotated over a 190$^{\circ}$ range in 1$^{\circ}$ steps, counting 30 seconds per angle. Background data were collected with 1/3 of the counting statistics at 50~K, 100~K, and 150~K over the same range. Data reduction and background subtraction were performed using the MJOLNIR \citep{MJOLNIR} software package.

\section{\label{sec:level1}Experimental results\protect\\}

\subsection{\label{sec:level2}Magnetic Susceptibility}

\begin{figure}[b]
\includegraphics[clip, trim=0.0cm 0.0cm 0.0cm 0.0cm ,width=8 cm]{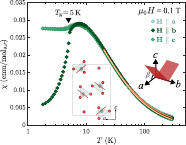}% Here is how to import EPS art
\caption{\label{fig:susceptibility} Main panel: magnetic susceptibility of \CRO single crystal in applied fields $\mo H$ = 0.1~T along the three crystallographic directions (symbols). The response along the $a$ and $c$ directions is almost indistinguishable in the plot. The solid lines indicate fits to the data as described in the text. Lower inset: a schematic view of the local ionic easy-planes as seen along the $b$ axis. Right inset: orientation of the two types of easy-planes relative the the crystallographic axes.}
\end{figure}

Bulk magnetic susceptibility curves ($\chi$), measured with a probing field of $\mo  H = 0.1$~T applied along each of the three crystallographic directions, are shown in Fig.~\ref{fig:susceptibility}. A typical anomaly characteristic of an antiferromagnetic ordering transition is observed at $T_{N}$ = 5~K. A Curie-Weiss fit in the range 50 - 300~K yields  an average Weiss temperature of $\theta_{CW}$ = -16~K, indicating some degree of frustration ($T_N$/$\theta_{CW}$ $\approx$ 3). The average effective moment is $\mu_{\mathrm{eff}}$ = 2.62~$\mub$, which is close to the $S$ = 1 spin-only value ($2\sqrt{S(S+1)}$ = 2.83~$\mub$), as expected for the low-symmetry distorted tetrahedral local environment, where angular momentum should be quenched.

Below the transition, the measured susceptibility shows a typical easy-$b$-axis behavior. Nevertheless, based on studies of the isostructural cobalt compounds, the underlying {\em local} anisotropy is likely to actually be of the easy-plane type \cite{Povarov2020, CCoC1}. The effective bulk easy-axis emerges at the intersection of the local easy-planes that are alternating for different ions \cite{Povarov2020}. By symmetry, the ionic easy planes are parallel to the crystallographic $b$ axis, but are alternately tilted in adjacent chains with respect to the $a$ axis. The tilt angle $\beta$ is not symmetry-constrained. This is illustrated in the inset in Fig.~\ref{fig:susceptibility}.
The single-ion anisotropy was assumed to be of the form $DS_z^2$. The parameters $D$ = 0.25~meV and the angle $\beta$ = 45$^\circ$ were determined from inelastic neutron scattering experiments (see below). With these fixed, components of the $g$-tensor were determined by simultaneously fitting the high temperature susceptibility data for different field orientations. Assuming $S$ = 1 degrees of freedom and an effective mean-field interaction constant $J_0$, yields an excellent fit in the temperature range $20$--$300$~K. The resulting parameters are $J_0$ = 2.13(3)~meV, with $g$-factors of $g_a$ = 1.89(4), $g_b$ = 1.88(4), and $g_c$ = 1.85(4). The results are indicated by the solid lines in Fig.~\ref{fig:susceptibility}.

\subsection{\label{sec:level2}Magnetization and Magnetic Torque}

\begin{figure}[t]
    \includegraphics[clip, trim=0.0cm 0.0cm 0.0cm 0.0cm ,width=8cm]{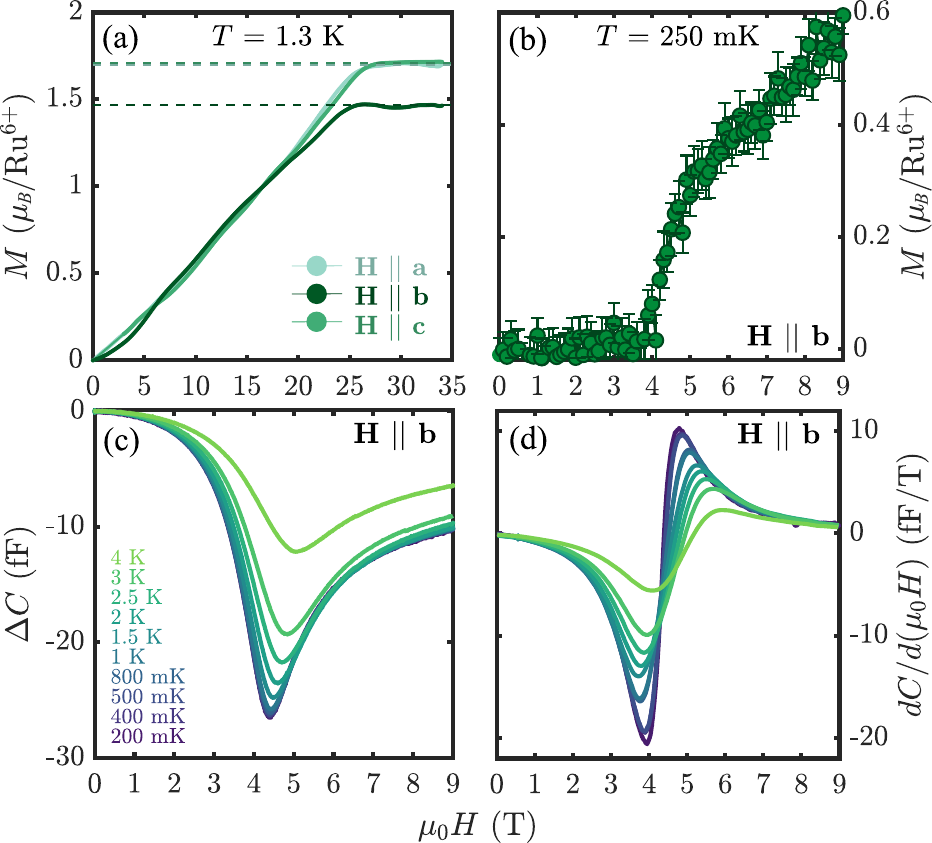}% Here is how to import EPS art
    \caption{\label{fig:magnetization} (a) High field magnetization per Ru$^{6+}$ measured at 1.3~K in \CRO for magnetic fields along the crystallographic $a$, $b$, and $c$ directions. Horizontal lines indicate the saturation magnetization for each curve. (b) Magnetization per Ru$^{6+}$ measured at 250 mK for a magnetic field along the crystallographic $b$ direction. Magnetic torque ($\Delta C$) (c) and its field derivative (d) measured at constant temperatures against magnetic field along the crystallographic $b$ direction. For all the datasets a capacitance reference value at zero field has been chosen and subtracted.}
\end{figure}

High field magnetization curves for the three field geometries up to 35~T at $T$ = 1.3~K are shown in Fig.~\ref{fig:magnetization}(a). Along the $a$ and $c$ axes, the magnetization increases linearly and saturates at $\mo H_{\mathrm{sat}} = 25.7$~T and $26.6$~T, with $M_{\mathrm{sat}} = 1.70$~$\mub$ and $1.71$~$\mub$, respectively. Along the $b$ axis, the magnetization is slightly suppressed at low fields and exhibits a weak anomaly around 4~T, before linearly increasing to saturation at $\mo H_{\mathrm{sat}}$ = 24.8~T and $M_{\mathrm{sat}} = 1.46$~$\mub$. Note that the $g$-factors extracted from susceptibility and  those estimated from the saturation magnetization are not entirely consistent. The former are almost perfectly isotropic, whereas among the latter $g_b$ is significantly lower than $g_a$ and $g_c$.

A low temperature magnetization measurement along the $b$ direction at 250~mK is shown in Fig.~\ref{fig:magnetization}(b). The anomalous behavior around 4~T is very apparent. The magnetization remains fully suppressed below it, then increases rapidly until 5~T, followed by a linear regime up to 9~T. The anomaly thus spans over a field range of almost 1~T, but is otherwise reminiscent of a spin-flop transition. That it involves spin re-orientation is also suggested by it showing a continuous anomaly in magnetic torque experiments, whereas a first-order transition would result in a discontinuity in torque and/or its derivative. Examples of a first-order spin-flop transition with a discontinuous response in torque measurements were observed in the related compound Cs$_2$CoBr$_4$ \cite{Povarov2020} as well as in chain-like compound Cs$_2$Cu$_2$Mo$_3$O$_{12}$ \cite{flavian}. The measured magnetic torque measurements and its field derivative for a field applied along the $b$ direction are shown in Fig.~\ref{fig:magnetization}(c) and (d), respectively. The data are presented as the change in measured capacitance $\Delta C$ = $C(H)$ - $C(H = 0~\mathrm{T})$.

\subsection{\label{sec:level2}Heat Capacity}

\begin{figure}[t]
\includegraphics[clip, trim=0.0cm 0.0cm 0cm 0.0cm ,width=8.5 cm]{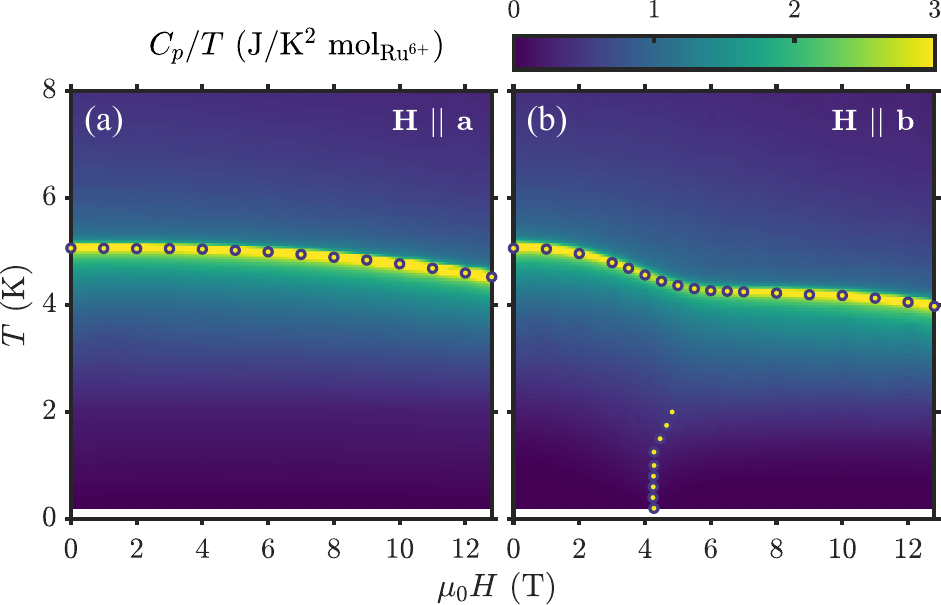}% Here is how to import EPS art
\caption{\label{fig:hc_ht} False-color plot of the temperature-scales specific heat measured in \CRO in a magnetic field applied along two crystallographic directions. Symbols indicate the positions of sharp lambda anomalies at the phase boundary and the additional low-temperature feature.}
\end{figure}

The magnetic phase diagram was mapped out through specific heat measurements with a field applied along the $a$ and $b$ directions, as shown in Fig.~\ref{fig:hc_ht}(a), (b), respectively.

\begin{figure}[t]
\includegraphics[clip, trim=0.0cm 0.0cm 0cm 0.0cm ,width=8 cm]{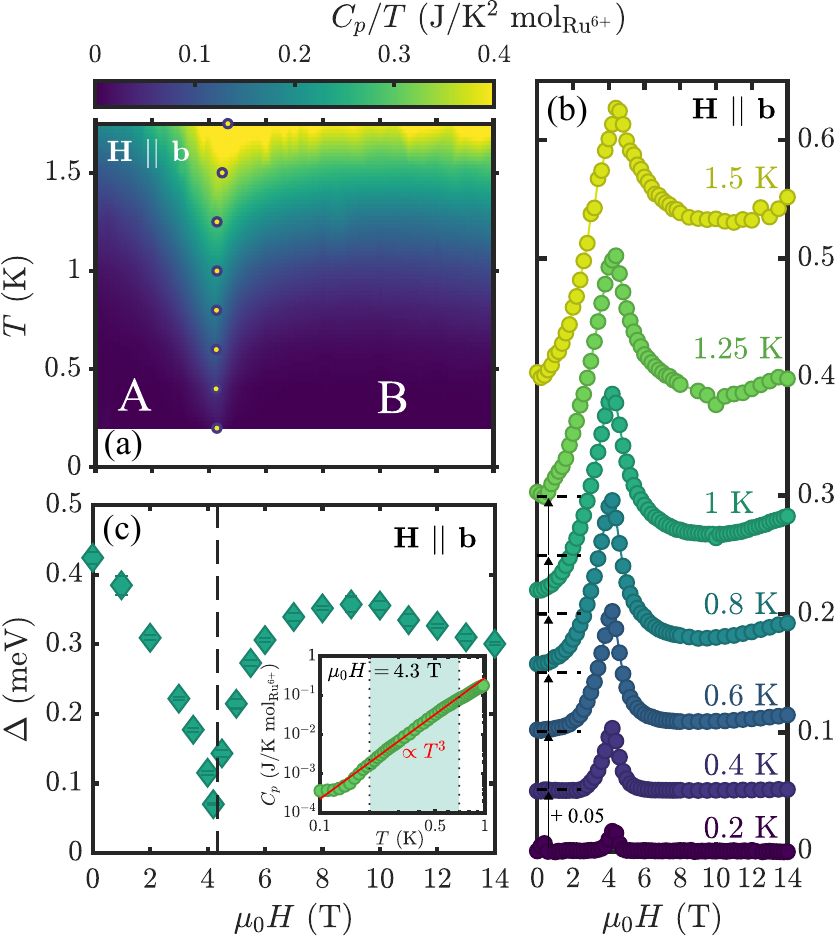}% Here is how to import EPS art
\caption{\label{fig:hc_lt_good} (a) Same as Fig.~\ref{fig:hc_ht} (b), zoomed in below 1.75~K and with enhanced color scale. (b) Specific heat measured at a constant temperature against applied magnetic field along the $b$ direction. Each curve is displaced with a constant offset of 0.05~J/K$^2$ mol$_\mathrm{Ru^{6+}}$ for visibility. (c) Activation gap $\Delta$  as a function of applied magnetic field along the $b$ direction. The dashed line indicates the critical field $\mo H_c\approx 4.3$~T. Inset: specific heat as function of temperature measured at $\mo H_c$. Note the logarithmic scale.}
\end{figure}

In zero field, the system exhibits a clear lambda anomaly at $\Tn $ = 5 K that we associate with the magnetic ordering transition. For $\bm{H} \| \bm{a}$ the phase boundary is largely insensitive to applied magnetic field in the studied range, decreasing only slightly to 4.5~K at 12.8~T. In all cases, below the transition temperature the heat capacity is exponentially suppressed, and no further anomalies are observed. In contrast, for $\bm{H} \| \bm{b}$ the ordering temperature dips between 3 and 6~T, then  gradually decreases to 4~K at 12.8~T. In this geometry a more subtle feature becomes apparent at lower temperatures as shown in  Fig.~\ref{fig:hc_lt_good}(a). Inside the ordered phase  a clear enhancement of the specific heat is seen towards the lowest temperatures at $H\sim 4$~T, delineating two distinct regions labeled A and B. In field scans at a constant temperature the anomaly manifests itself as distinct peaks, as shown in Fig.~\ref{fig:hc_lt_good}(b). The weaker feature seen in the curve at 0.2 K for $\mo H$ $<$ 1~T is attributed to a residual Schottky anomaly from the silver sample mount.

In both regions A and B, specific heat shows an activated behavior as a function of temperature.
The activation gap $\Delta$ was estimated using simple exponential fits to the low-temperature data: $C \propto \exp[-{\Delta}/{k_bT}]$. The thus determined field dependence of the gap is shown in Fig.~\ref{fig:hc_lt_good}(c). To within experimental accuracy the gap closes at $\mo H_c=4.3$~T. 

The softening and re-opening of the gap and the continuous nature of the phase transition
that occurs at $T\rightarrow 0$ is a clear sign of a quantum phase transition and critical point. Criticality is also apparent in  the characteristic scaling of specific heat, as shown in the inset of  Fig.~\ref{fig:hc_lt_good}(c). Between 0.2 K and 0.7 K, the data are consistent with $C\propto T^3$ behavior indicative of $z=1$ criticality in this moderately frustrated three-dimensionally coupled spin system (see below).
Any deviations at lower temperature may be a result of a mismatch
between the the values of measurement- and actual
critical fields. Future dedicated studies of scaling behavior may shed more light on this issue.

\subsection{\label{sec:Diffraction}Neutron Diffraction}

\begin{figure}[h]
\includegraphics[clip, trim=0.0cm 0.0cm 0.0cm 0.0cm ,width=6 cm]{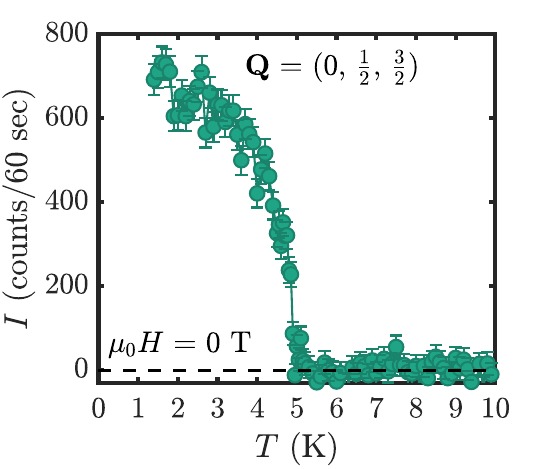}% Here is how to import EPS art
\caption{\label{fig:OP} Temperature dependence of $\omega$-integrated intensity of the $(0, 1/2, 3/2)$ magnetic Bragg reflection in zero magnetic field. A constant background measured above $T=5$~K has been subtracted.}
\end{figure}

\begin{table}[b!]%The best place to locate the table environment is directly after its first reference in text
\caption{\label{tab:diff}%
The first column indicate the ($h$,$k$,$l$) indices of four measured magnetic reflections. The second and third column show the measured $\omega$-integrated intensity normalized by the Lorentz factor ($I/L$) and calculated magnetic structure factors ($F_{\mathrm{calc}}^2$).}
\begin{ruledtabular}
\begin{tabular}{c|c|c}
\textrm{Reflection ($h$,$k$,$l$)}&
\textrm{$I/L$ ($\times 10^3$)(a.u.)}&
\textrm{$F^2_{\mathrm{calc}}$ ($\times 10^3$)(a.u.)}\\
\colrule
(0,1/2,1/2) & 0.3(1) & 0.2\\
(0,1/2,3/2) & 2.0(1) & 2.0\\
(1,1/2,1/2) & 1.0(1) & 1.1\\
(2,1/2,3/2) & 0.6(1) & 0.8\\ 
\end{tabular}
\end{ruledtabular}
\end{table}

Magnetic Bragg reflections appear in neutron diffraction experiments at temperatures below $\Tn$ and correspond to a propagation vector $\bm{k} = (0, 1/2, 1/2)$. The temperature dependence of the $(0, 1/2, 3/2)$ $\omega$-integrated intensity measured in \CRO is shown in Fig.~\ref{fig:OP}. Unfortunately, the sample used in the measurement contained three crystallographic domains. The domains share a common $a$ axis, but are rotated around it by approximately $60^{\circ}$ relative to one another. The most likely cause is a suspected hexagonal-to-orthorhombic phase transition during crystal growth at 634~$^{\circ}$C \cite{structrans}. Most magnetic reflections overlap with nuclear ones from other domains, which hindered a proper magnetic structure refinement. However, certain peaks, such as  $(0, 1/2, 3/2)$ for instance, are either not affected or overlap with very weak nuclear peaks. Because of this, only four magnetic peaks (listed in Table~\ref{tab:diff}) could be unambiguously indexed. The effect of weak nuclear peak contamination was mitigated by subtracting the intensity at 10 K. This results in the correct relative intensities. Their $\omega$-integrated intensities, normalized by the Lorentz factor $L$ and a scale factor obtained from nuclear peaks, is shown in the second column of Table~\ref{tab:diff}. These are consistent with the calculated structure factors ($F^2_{\mathrm{calc}}$) of a colinear Néel structure, with the magnetic moments parallel to the $b$ axis, as in Cs$_2$CoBr$_4$ \cite{Facheris2022}. Unfortunately, the magnetic form factor for Ru$^{6+}$ is not tabulated in Ref.~\cite{formfac}, which is required to calculate $F^2_{\mathrm{calc}}$. Instead we utilized the same approach as in Refs.~\cite{formfac2, NRO}, approximating it by the known form factor of the lighter 4$d$ element Zr$^{+}$. The only fitting parameter used here is the moment-size of Ru$^{6+}$ resulting in 1.6(2)~$\mu_B$, which is consistent with the full expected moment-size of $g_bS$ = 1.5~$\mu_B$, where the errorbars have been appropriately enlarged due to the systematic error introduced by using a different magnetic form factor.

\subsection{\label{sec:INS}Neutron Spectroscopy}

\begin{figure}[b]
\includegraphics[clip, trim=0.0cm 0.0cm 0.0cm 0.0cm ,width=8 cm]{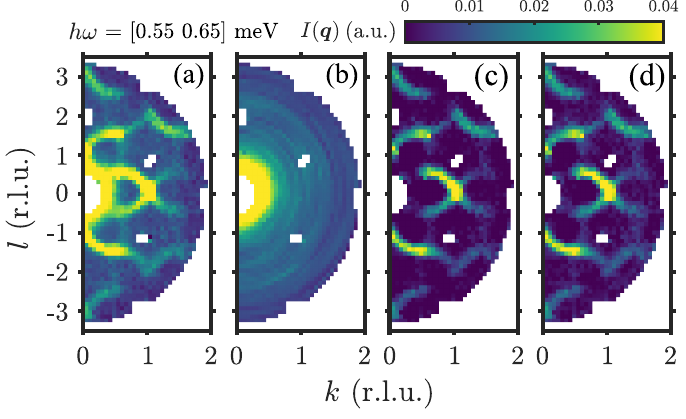}% Here is how to import EPS art
\caption{\label{fig:Bgd_procedure} Exemplary constant energy cuts at $\hbar \omega$ = 0.6~meV, where (a) is the raw measured data, (b) the smoothed extrapolated background, (c) the background subtracted data and (d) with an additional constant offset. The white areas mask regions affected by Currat-Axe spurious scattering.}
\end{figure}

Inelastic neutron scattering measurements were performed in zero magnetic field at $T$ = 1.6~K. The sample was aligned with the $(0,k,l)$ crystallographic plane in the scattering plane of the spectrometer. The non-magnetic background was subtracted using datasets at $T$ = 50, 100 and 150~K. Exemplary constant energy cuts in Fig.~\ref{fig:Bgd_procedure} at $\hbar \omega$ = 0.6~meV illustrate the background subtraction procedure. The raw measured data is shown in Fig.~\ref{fig:Bgd_procedure}(a). We assume that the total scattering intensity $I$ can be decomposed into
magnetic scattering, a temperature-independent background due to incoherent scattering in the sample and scattering in the sample environment, etc., and phonon scattering that scales with temperature as the Bose factor:
\begin{equation}
\begin{split}
I(\textbf{q}, \omega, T) &= I_{\text{mag}}(\bm{q}, \omega, T) + I_{\text{indep}}(\bm{q}, \omega) \\
&\quad + I_{\text{phonon}}(\bm{q}, \omega) \left(1-e^{-\frac{\hbar\omega}{k_b T}}\right)^{-1}
\end{split}
\end{equation}

We further assumed that the magnetic part was negligible except at base temperature, and that the phonons and incoherent background simply scale as a function of temperature. The latter two contributions where extrapolated from high temperatures point by point. The resulting background was then smoothed with a Gaussian filter with widths of 10$^{\circ}$ in the rotation angle and roughly 0.6$^{\circ}$ in 2$\theta$ and not smoothed in energy [Fig.~\ref{fig:Bgd_procedure}(b)]. This was then point by point subtracted from the raw spectrum [Fig.~\ref{fig:Bgd_procedure}(c)]. Deviations from both assumptions that magnetic scattering is negligible at high temperatures and a simple scaling of the background components results in a very slight over-subtraction. This effect was mitigated by adding a small constant (energy and momentum transfer-independent) offset of +0.002 a.u. to the resulting intensity. The final result is illustrated in Fig.~\ref{fig:Bgd_procedure}(d).

\begin{figure}[!t]
\includegraphics[clip, trim=0.0cm 0.0cm 0.0cm 0.0cm ,width=8cm]{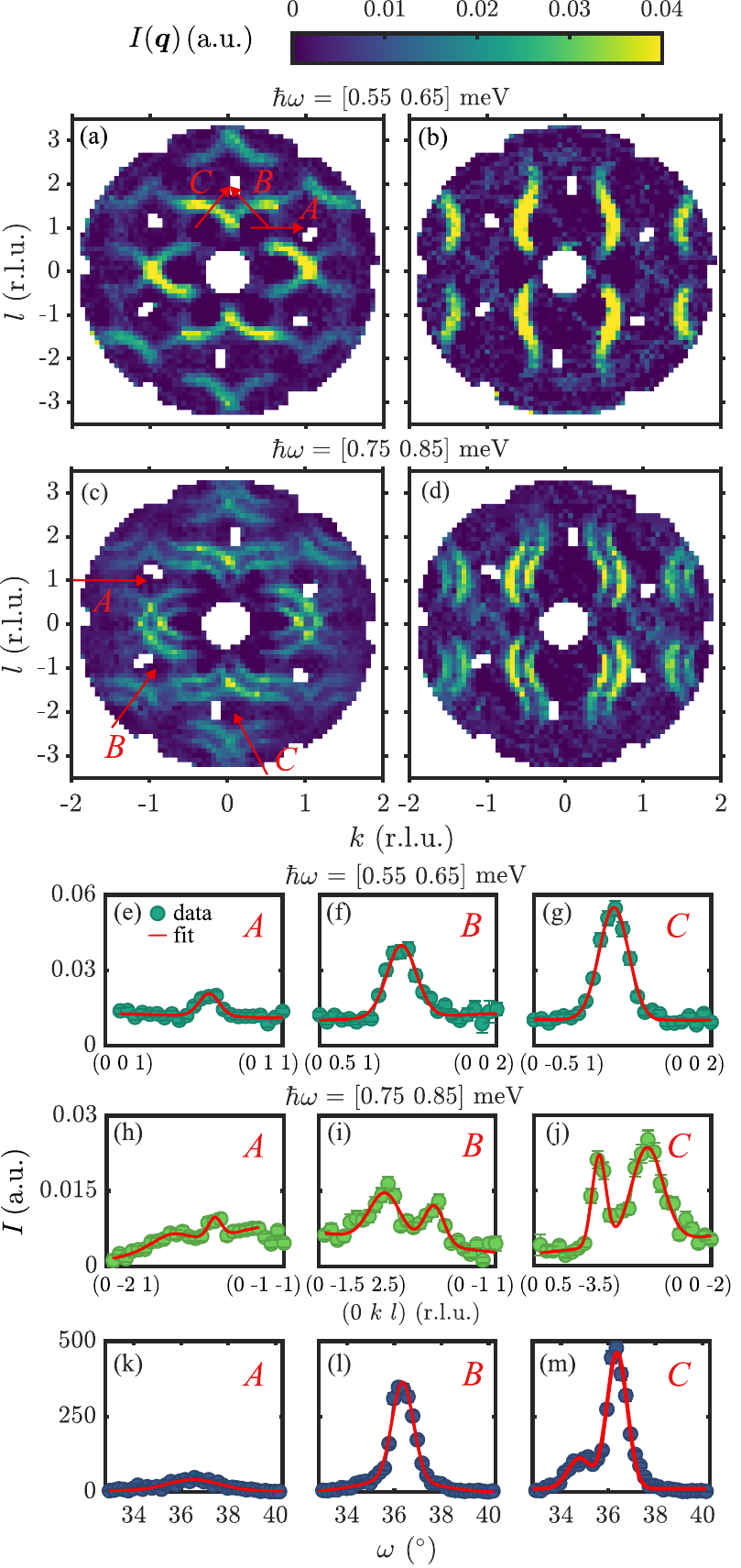}% Here is how to import EPS art
\caption{\label{fig:detwinning} False color plot of the inelastic neutron intensity before (a), (c) and after (b), (d) applying the domain-separating procedure described in the text for $\hbar \omega$ = 0.6 and 0.8~meV, respectively. The intensity scale in (b), (d) represents the summed spectral weight of the tree domains. The actual data were collected in a $190^\circ$ sample rotation range and symmetrized to fill the complete $360^\circ$ angle for visual purposes. Red lines indicate $\bm{q}$-cuts, with labeling corresponding to the intensity plots in (e)-(g) and (h)-(j) for $\hbar \omega$ = 0.6 and 0.8~meV, respectively. In the latter the intensities were integrated perpendicular to the cut trajectory in the range $\pm$~0.15 r.l.u. in the $k$-$l$ plane, and $\pm$~0.05~meV in energy. Rocking curves of nuclear peak (3,0,1) from the three different domains are shown in (k)-(m).}
\end{figure}

\begin{figure*}[t]
\includegraphics[clip, trim=0.0cm 0.0cm 0.0cm 0.0cm ,width=\textwidth]{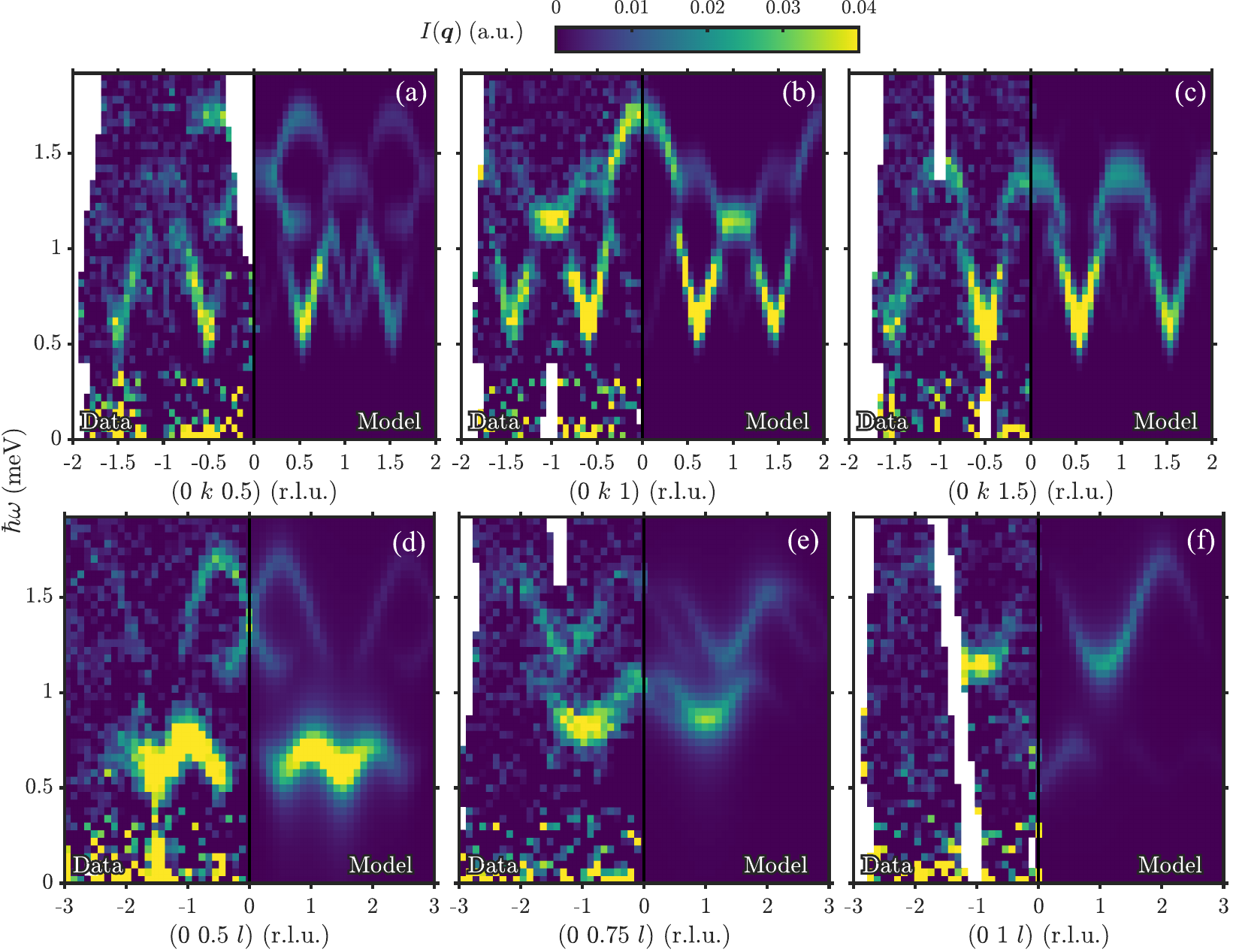}% Here is how to import EPS art
\caption{\label{fig:QE} On the left side of each plot are false color representations of energy-momentum projections of single-domain $\mathcal{S}(\bm{q},\omega)$ measured in \CRO at $T$ = 1.6~K. The data are integrated perpendicular to the scan direction in the $k$-$l$ plane in the range $\pm0.1$~r.l.u. along $k$ and in the range $\pm0.1$~r.l.u. along $l$. On the right side of each plot are $SU$(3) spin wave theory calculation based on the minimal parameters listed in Table \ref{tab:interactions}.}
\end{figure*}

As in the diffraction experiment, the sample used for inelastic studies consisted of three crystallographic domains, rotated relative to one another by approximately $60^{\circ}$ around the $a$ axis.
Correspondingly, the measured spectrum consists of three superimposed contributions. To illustrate this, a representative constant energy cut at an energy transfer of $\hbar \omega = 0.6$~meV and 0.8 meV are shown in Fig.~\ref{fig:detwinning}(a),(c). The unequal size of the three domains makes it possible to recover the intrinsic single-domain scattering pattern. We assume that the intensity $I(\bm{q}, \omega)$ measured at a given momentum and energy transfer receives contributions from all three domains proportionately to their relative masses $A$, $B$, and $C$. The total intensity can thus be written as a linear combination of three single-domain cross sections $\mathcal{S}(\bm{q},\omega)$ contributed by domain $A$, $\mathcal{S}(\mathfrak{R}_+\,\bm{q},\omega)$ from domain $B$, and $\mathcal{S}(\mathfrak{R}_-\,\bm{q},\omega)$ from domain $C$. Here $\mathfrak{R}_\pm$ denotes rotation matrices around the $a$-axis by $\pm60^{\circ}$  degrees, respectively. Applying this argument to the measured intensities $I(\mathfrak{R}_+\,\bm{q},\omega)$, and $I(\mathfrak{R}_-\,\bm{q},\omega)$ yields a set of three linear equations:

\begin{equation}
\begin{bmatrix}
I(\bm{q}, \omega)  \\
I(\mathfrak{R}_+\,\bm{q}, \omega)  \\
I(\mathfrak{R}_-\,\bm{q}, \omega)
\end{bmatrix}
=
\begin{bmatrix}
A & B & C \\
C & A & B \\
B & C & A
\end{bmatrix}
\begin{bmatrix}
\mathcal{S}(\bm{q}, \omega)  \\
\mathcal{S}(\mathfrak{R}_+\,\bm{q}, \omega)  \\
\mathcal{S}(\mathfrak{R}_-\,\bm{q}, \omega)
\end{bmatrix}
\label{eq:matrix}
\end{equation}

As long as the matrix is not singular, the system of linear equations can be solved to extract the scattering cross section $\mathcal{S}(\bm{q}, \omega)$ of a single domain without any additional assumptions. In our case, the mass ratios were extracted in two different ways. First, by examining inelastic signal in constant-energy cuts that clearly contain features produced by particular domains. Specifically, we chose cuts at $\hbar \omega$ = 0.6~meV and 0.8~meV along the red arrows labeled $A$, $B$, and $C$ in Fig.~\ref{fig:detwinning}(a), (c). The corresponding intensity profiles are shown in Fig.~\ref{fig:detwinning}(e)-(g) and (h)-(j), respectively (no background subtracted), each showing a prominent peak that we assume to originate from a single domain. Additionally, we have measured the nuclear peak (3,0,1) for each domain at room temperature [Fig.~\ref{fig:detwinning}(k)-(m)]. Following the rotation matrices, for the other domains, this peak will overlap with the magnetic peak positions (3,1/2,1/2) and (3,-1/2,1/2) as indexed in the original domain. Since this was measured at room temperature, and no magnetic signal is present, this intensity can be solely attributed to a single domain. In peak $C$ [Fig.~\ref{fig:detwinning} (m)] a shoulder is observed next to the peak, which is attributed to mosaic. All set of peaks were analyzed using Gaussian profiles.  From the fitted intensities, the extracted mass ratios were consistent and the averaged values $A = 0.09(1)$, $B = 0.35(2)$ and, $C = 0.56(2)$ were determined. The corresponding domain matrix is far from singular. Solving the linear equations yields the single-domain cross section $\mathcal{S}(\bm{q},\omega)$ at the same energy transfer, as shown in Fig.~\ref{fig:detwinning}(b). All neutron spectra discussed below are after applying this procedure and with the background subtracted as described above.

\begin{figure*}[t]
\includegraphics[clip, trim=0.0cm 0.0cm 0.0cm 0.0cm ,width=18cm]{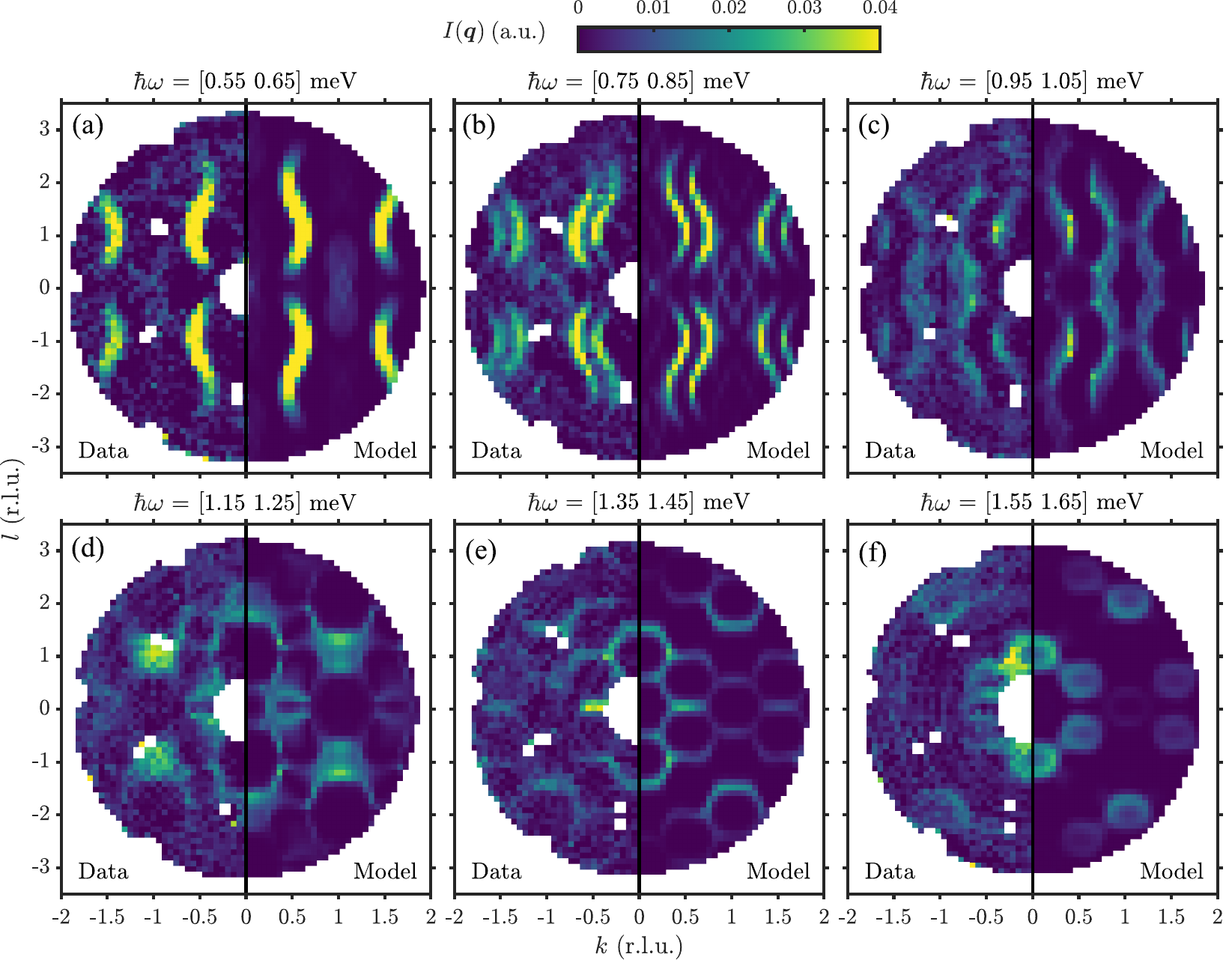}% Here is how to import EPS art
\caption{\label{fig:E} On the left side of each plot are false color representations of constant energy cuts of single-domain $\mathcal{S}(\bm{q},\omega)$ measured in \CRO at $T$ = 1.6~K. The integration range in energy is $\pm$~0.05~meV. On the right side of each plot are $SU$(3) spin wave theory calculations based on the minimal parameters listed in Table \ref{tab:interactions}.}
\end{figure*}

Representative energy-momentum projections and constant energy slices of the single-domain spectra measured in \CRO are shown in Fig.~\ref{fig:QE} and Fig.~\ref{fig:E}, respectively.  These data reveal a complex spectrum of resolution-limited modes dispersing along both the $k$ and $l$ directions. Based on Fig.~\ref{fig:QE}, the spectrum can be broadly divided into two separate but possibly hybridized magnon branches: the first extending to approximately  1.1~meV, and the second in the range between 1.1 to 1.6~meV. No significant spectral weight is observed above 1.6~meV. The lower branch exhibits a gap of $\Delta \approx$ 0.55~meV, the dispersion minimum coinciding with the positions of the magnetic Bragg peaks.

\subsection{\label{sec:Analysis}INS Data Analysis and Model Hamiltonian}

With the goal to determine a minimal model spin Hamiltonian for \CRO, the inelastic neutron scattering data were analyzed using $SU$(3) spin wave theory (SWT) computations performed using the SUNNY.jl software package \cite{sunny}. In this we followed an approach similar to that previously used for \CCB~\cite{Facheris2024}. Restricting interactions to just the 5 nearest-neighbor bonds already allows for as many as 32 anisotropic exchange parameters that are not constrained by symmetry. To reduce the number of free parameters we assumed that all interactions between the $S=1$ Ru$^{6+}$ spins are fully isotropic (Heisenberg), and that magnetic anisotropy is entirely due to the single-ion term. The model used included exchange constants up to the eighth nearest neighbor. The remaining two parameters were as discussed above: the single-ion anisotropy constant $D$ and the tilt angle $\beta$ of the local anisotropy planes relative to the crystallographic $a$ axis. Again, just like in Section~\ref{sec:Diffraction}, since the magnetic form factor for Ru$^{6+}$ is unknown, it is approximated by the known form factor of the lighter 4$d$ element Zr$^{+}$ in a different oxidation state.

Using this model with only the first 5 exchange constants $J_1$--$J_5$, as previously done for \CCB \cite{Facheris2024}, can produce a reasonable good fit to the spectra measured in \CRO. By trial and error we found that an almost perfect fit can be obtained by also including one particular coupling from the next shell, namely $J_8$. At the same time, $J_5$ was dropped from the model since including it did little to improve agreement. As was the case for \CCB~\cite{Facheris2024}, the parameters $J_1$ from $J_3$ are practically impossible to distinguish due to their similar geometric roles in the magnetic lattice. Somewhat arbitrarily we set $J_3$ to zero, and used only the  $J_1$ coupling, which corresponds to a slightly shorter bond length.
Using reasonable starting values, the remaining six free parameters were refined by fitting excitation energies determined for 160 constant-$\bm{q}$ energy scans extracted from the data. The final optimized parameters and standard deviations are summarized in Table \ref{tab:interactions}. Here we also indicate the corresponding bond length and whether or not the interaction connects spins with parallel or almost orthogonal single-ion anisotropy planes. The simulated spectra are shown alongside the data in Figs.~\ref{fig:QE} and \ref{fig:E}. Despite all the assumptions made, the model captures the key features remarkably well.

\begin{table}[t]%The best place to locate the table environment is directly after its first reference in text
\caption{\label{tab:interactions}%
Parameters of a minimal model Heisenberg Hamiltonian for \CRO. The third column indicates whether the interaction connects parallel or perpendicular anisotropy planes}
\begin{ruledtabular}
\begin{tabular}{c|c|c|c}
\textrm{ }&
\textrm{Bond length (\AA)}&
\textrm{Anisotropy}&
\textrm{Value}\\
\colrule
$J_1$ & 5.29 & $\parallel$ & 0.20(2) meV\\
$J_2$ & 5.80 & $\perp$ & 0.44(5) meV\\
$J_4$ & 6.22 & $\parallel$ & 0.22(3) meV\\
$J_8$ & 8.69 & $\perp$ & -0.016(4) meV\\
$D$    & -- & -- & 0.25(2) meV\\
$\beta$ & -- & -- & 45(5)$^{\circ}$\\
\end{tabular}
\end{ruledtabular}
\end{table}

\section{\label{sec:level1}Discussion\protect\\}

\subsection{\label{sec:compare}Comparison with Cs$_2MX_4$ family}

\begin{figure}[b]
\includegraphics[clip, trim=0.0cm 0.0cm 0.0cm 0.0cm ,width=9cm]{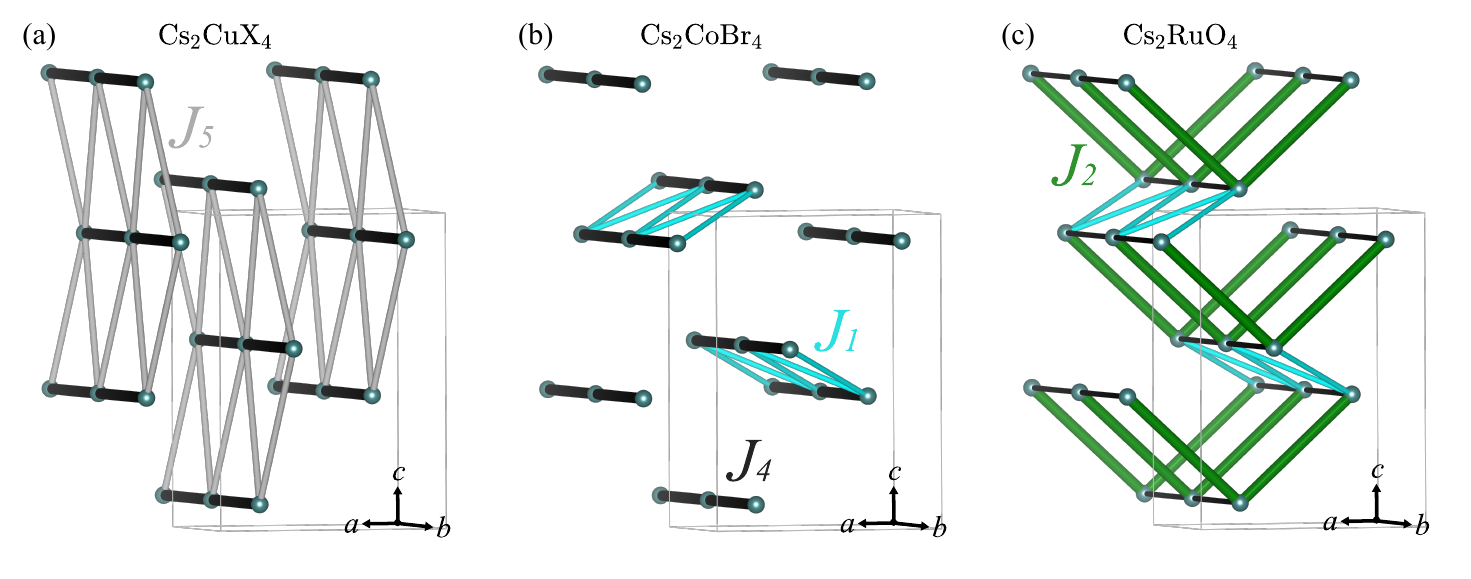}% Here is how to import EPS art
\caption{\label{fig:exchanges} Schematic overview of the strongest magnetic interactions in (a) Cs$_2$CuX$_4$ (X = Cl, Br), (b) Cs$_2$CoBr$_4$, and (c) Cs$_2$RuO$_4$. Less significant interactions are left out for visibility.}
\end{figure}

This work sheds new light on the interaction hierarchy in the Cs$_2MX_4$ family of frustrated magnets. Distorted triangular-lattice connectivity (spanned by $J_4$ and $J_5$) is known to dominate in the copper- and halogen-based magnets \cite{CCC2, CCB1}, as schematically illustrated in Fig.~\ref{fig:exchanges}(a). Cs$_2$CoBr$_4$ has a more one-dimensional zig-zag ladder type character spanned by $J_4$ and $J_1$ [Fig.~\ref{fig:exchanges}(b)]. In contrast, \CRO features a much more pronounced 3D interaction network \cite{Facheris2024} [Fig.~\ref{fig:exchanges}(c)]. This material can be seen as composed rectangular lattices spanned by $J_2$ and $J_4$. There is no geometric frustration in {\em interactions} in these layers. However, the strongest coupling $J_2$ connects magnetic ions with orthogonal anisotropy planes resulting in {\em strong frustration of anisotropy}. As discussed below, this has important implications for the phase behavior.
The coupling between the rectangular planes is provided by the zig-zag $J_1$ interaction. It is of similar size as $J_4$ but entirely frustrated. The fact that \CRO has such a distinct interaction structure compared to other materials in the family is perhaps not surprising. After all, the magnetic ion is different and the super-exchange between them is mediated by O$^{2-}$ anions rather than halogens.

\subsection{\label{sec:compare}Continuous spin-flop-like transition and QCP}

\begin{figure*}[t]
\includegraphics[clip, trim=0.0cm 0.0cm 0.0cm 0.0cm ,width=\textwidth]{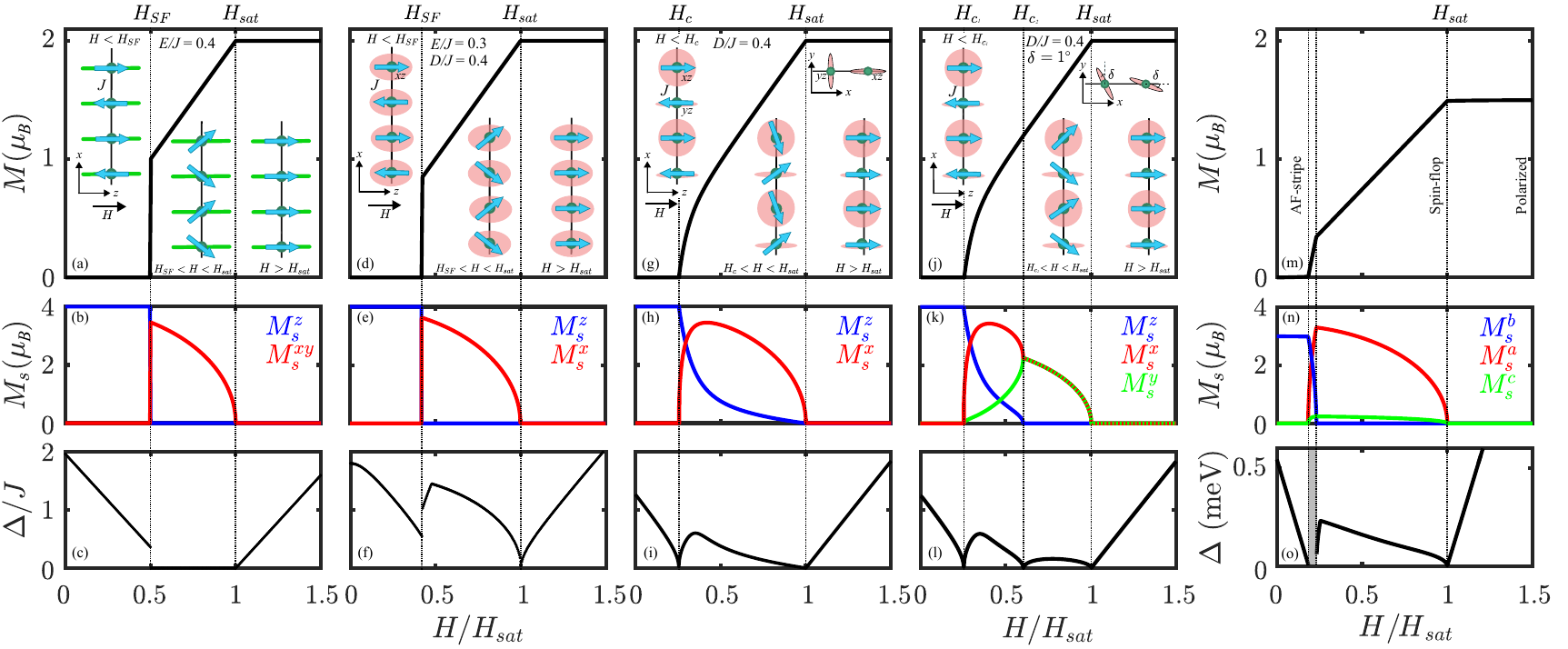}% Here is how to import EPS art
\caption{\label{fig:sfcm} Classical simulations of a Heisenberg chain with four different scenarios for single-ion anisotropy: easy axis anisotropy (a-c), uniform XYZ anisotropy (d-f), alternating orthogonal anisotropy easy-planes (g-i), and alternating non-orthogonal anisotropy planes (j-l). The calculated quantities are magnetization ($M$), components of the staggered magnetization ($M_s^z$ blue, $M_s^x$ red, $M_s^y$ green), and the excitation gap ($\Delta/J$) as a function of applied magnetic field applied along the $z$ axis (rows from top to bottom). The identified phase transitions are indicated with a dotted line. An illustration of the corresponding  ground states is shown where appropriate. Additionally SU(3) simulations of the same quantities of the full model Hamiltonian with $\beta$ = 44$^{\circ}$ are shown in (m-o). In (n), components of the staggered magnetization correspond to the crystallographic axes ($M_s^b$ blue, $M_s^a$ red, $M_s^z$ green).}
\end{figure*}

Of particular interest is  the spin-flop-like transition accompanied by a QCP that we observe in \CRO in magnetic fields applied along the $b$ axis. An archetypal spin flop like the one seen in, for instance, GdAlO$_3$ \cite{PhysRev.173.574}, connects phases with two distinct order parameters: longitudinal staggered magnetization at low fields and transverse staggered magnetization at high fields. Within Landau's paradigm, transitions between phases with different spontaneously broken symmetries are necessarily discontinuous. The one seen in \CRO is clearly continuous and is associated with a quantum critical point at $T\rightarrow 0$, $H\rightarrow H_c$. It turns out that the key reason for this unusual behavior is the strong anisotropy frustration  mentioned above. The underlying physics is fully captured by a simple toy model: classical antiferromagnetic spin chains with alternating orthogonal anisotropy planes corresponding to those spanned by the dominant interaction  $J_2$ in \CRO.

In order to demonstrate this, we investigated the field dependencies of the uniform magnetization ($M$), staggered magnetization ($\bm{M}_s$) and excitation gap ($\Delta/J$) in this simple construct.  The ground state was obtained by numerically minimizing the classical energy. The gap was computed using SUNNY.jl within linear spin wave theory. Results for several spin chain models are compared in Fig.~\ref{fig:sfcm}. Each column corresponds to a particular anisotropy configuration. The first two columns represent a conventional spin flop transition with no frustration of anisotropy. The corresponding Hamiltonian is:

\begin{align}
\hat{\mathcal{H}} 
&= \sum_i J \bm{S}_i \bm{S}_{i+1} - g \mu_\text{B} \mu_0 H \sum_i S_i^z \nonumber \\
&\quad + \sum_i \left[ D (S_i^y)^2 - E (S_i^z)^2 \right]   
\end{align}

Here $J>0$ is the Heisenberg exchange constant, $E>0$ and $D>0$ are single-ion anisotropy parameters corresponding to easy axis  (along $z$) and easy-plane (in the $xz$ plane) terms, respectively, and $H$ is the strength of  magnetic field applied in the $z$ direction.

The first column of Fig.~\ref{fig:sfcm} corresponds to the textbook spin-flop transition: an antiferromagnetic chain with easy-axis anisotropy ($E/J$ = 0.3 and $D/J$ = 0). In zero field, the system exhibits a gapped, colinear Néel ground state with all spins along $z$. This state remains unchanged by applied field all the way to the transition at $H_\text{SF}$. At that point the magnetization shows a discontinuous jump [Fig.~\ref{fig:sfcm}(a)]. The staggered magnetization along $z$ abruptly disappears and is replaced by antiferromagnetic order of transverse spin components [Fig.~\ref{fig:sfcm}(b)]. The spectral gap abruptly drops to zero. The spin flop state is gapless due to a continuous SO(2) rotational symmetry around the easy axis [Fig.~\ref{fig:sfcm}(c)]. A continuous transition to a fully polarized paramagnetic state happens at some large field $H_\text{sat}$.

In the second column, we introduce an additional uniform easy-plane anisotropy ($E/J$ = 0.3 and $D/J$ = 0.4). The overall spin-flop transition remains qualitatively similar, marked by discontinuous behavior in both uniform and staggered magnetization [Fig.~\ref{fig:sfcm}(d,e)]. The only significant difference is that the spin flop phase is now also gapped, due to lack of rotational symmetry [Fig.~\ref{fig:sfcm}(f)]. The spin flop phase only spontaneously breaks a $\mathbb{Z}_2$ symmetry of the Hamiltonian.

The effect of frustration by alternating single-ion anisotropy planes is borne out in the third column. This scenario is closest to the situation in \CRO, where the anisotropy planes of consecutive spins are almost orthogonal. The corresponding Hamiltonian is:

\begin{align}
\hat{\mathcal{H}} 
&= \sum_i J \bm{S}_i \bm{S}_{i+1} - g \mu_\text{B} \mu_0 H \sum_i S_i^z \nonumber \\
&+ D\sum_{i-\text{odd}}(S_i^y)^2 + D\sum_{i-\text{even}}(S_i^x)^2  
\end{align}

In this case, the $z$-axis is equivalent to the $b$-axis in our material, in which the crossed anisotropy planes generate an effective easy-axis. In \CRO, the successive anisotropy planes are tilted in opposite directions with an angle $\beta$ with respect to the $a$ axis. As they are near orthogonal ($\beta$ = 45$^{\circ}$), the chain-$x$-axis is chosen such that the anisotropy planes are either parallel or perpendicular to it [see top right inset of Fig.~\ref{fig:sfcm}(g)]. This corresponds to the $x$-axis running at 45 degrees with respect to the $a$ axis in the material. This allows for clearer illustrations of the ground states. The tilt of an anisotropy plane $\delta$ in the model, corresponds to a tilt of $\beta$ = 45 - $\delta$ in the material [see top right inset of Fig.~\ref{fig:sfcm}(j)].

The simulation in the third column corresponds to $D/J = 0.4$, with orthogonal anisotropy planes ($\delta$ = 0$^{\circ}$). In zero field, we recover a conventional Néel state with all spins along $z$, within their respective easy-planes.
This state persists to the transition field $H_c$  as in a conventional spin flop. In contrast to the two above scenarios, at and beyond the transition the uniform magnetization [Fig.~\ref{fig:sfcm}(g)] evolves continuously. The high-field phase is not quite the conventional spin-flop state. All spin remain co-planar with the applied field. They do cant to form a transverse staggered magnetization, but the canting of {\em all} spins occurs in the {\em one} of the two anisotropy planes, spontaneously chosen. The magnitude of the canting angle therefore alternates along the chains. As a result, the longitudinal staggered magnetization persists beyond the transition [blue line in Fig. ~\ref{fig:sfcm}(h)]. There is indeed a new order parameter, the transverse staggered magnetization, but it grows continuously starting at the transition point [red line in Fig.~\ref{fig:sfcm}(h)]. As it does not replace the longitudinal staggered magnetization but appears alongside it, a continuous transition is not prohibited by Landau theory. The spectral gap continuously goes to zero at the transition point and continuously re-opens beyond it, consistent with a QCP [Fig.~\ref{fig:sfcm}(i)]. While the Hamiltonian does not have rotational symmetry around the $z$ axis, it does allow for four equivalent ordered phases of this type, with spins canting left or right, parallel to one or the other family of anisotropy planes. The transition is therefore of the 4-state vector-Potts universality class. The unconventional state persists all the way to the continuous saturation transition.

The near-orthogonality of anisotropy planes in \CRO is accidental, but a similar transition occurs also for $\beta$ deviating from $45^{\circ}$. Simulations for $D/J = 0.4$ and $\delta$ = 1$^{\circ}$ are shown in the last column of Fig.~\ref{fig:sfcm}. The transition at $H_{c1}$ occurs similarly as in the highly-symmetric case, but beyond it the spins are no longer co-planar with the applied field. Interestingly, there is now a {\em second} continuous transition at $H_{c2}<H_\text{sat}$. Here it is the longitudinal staggered magnetization that disappears continuously, even as the transverse one persists across the transition. The high field state is a conventional spin flop phase. Here, the spins are canted, as in a conventional spin-flop and form equal angles with both families of anisotropy planes. The transitions at $H_c$ and $H_{c2}$ are in the same universality class. 

The full 3D model Hamiltonian obtained from INS is evidently more complicated than the considered simple chain model. However, we have verified using SUNNY.jl, through SU(3) calculations of  magnetization, staggered magnetization, and the excitation gap, that the full Hamiltonian results in a similar continuous spin flop-like transition accompanied by a QCP. The simulations have now been performed for a applied magnetic field along the crystallographic $b$ direction.

We observe a similar rapid continuous increase in uniform magnetization [Fig.~\ref{fig:sfcm}(m)], however, an additional transition is visible where the slope changes. The staggered magnetization [Fig.~\ref{fig:sfcm}(n)] reveals that an intermediate phase emerges with both longitudonal and transverse staggered magnetization components in a narrow window, just like in the previous case. After this, it continuously evolves into a conventional spin-flop phase. Due to the competing easy-planes and antiferromagnetic interactions $J_2$ the spins will not flop exactly onto the easy-planes, but find a compromise, mostly along either the $a$ or $c$ direction which are accidentally degenerate if $\beta$ = 45$^{\circ}$. To lift the degeneracy of this 'special' angle, we have slightly tilted the angle to $\beta$ = 44$^{\circ}$, so the $a$ direction is favored. This represents a more 'general' case which is justified by the substantial errorbar in the determination of $\beta$. However, the behaviour of $M$ and $M_s$ over the transitions, as well as the width of the intermediate state have been verified to remain similar in both cases. In this case, the intermediate state is thus likely narrowed due to the 3D interactions besides our simple chain model, rather than by slightly changing the angle $\beta$ as seen in Fig.~\ref{fig:sfcm}(h) compared to (k). The ground state obviously needs to be further investigated through a neutron diffraction experiment. Lastly, in Fig.~\ref{fig:sfcm}(n) we see a closing and re-opening of the gap, however, near the critical point in the intermediate phase the gap simulation is unstable [gray shaded area in Fig.~\ref{fig:sfcm}(n)]. Due to the two transitions it could be that the excitation gap slightly opens again, having two QCP's in close vicinity. The exact behaviour of the gap should be further investigated through a in-field spectroscopy experiment.

The QCP in \CRO is distinct from those in typical magnetic quantum phase transitions, such as that in CoNb$_2$O$_6$ \cite{CoNbO} in that it happens {\em inside} the magnetically ordered state, rather that separating ordered and disordered phases. The key issue is that the "background" {\em longitudinal} order in \CRO is separate from the order-disorder transition in the {\em transverse} channel. Just like the presence of {\em crystallographic} long range order in CoNb$_2$O$_6$ is entirely separate from the {\em magnetic} order-disorder transition.

\section{\label{sec:level1}Conclusion\protect\\}
The $S=1$ compound \CRO is characterized by a strong frustration of magnetic anisotropy. This feature qualitatively accounts for the observed unusual ``continuous spin-flop transition'' and quantum critical point inside the magnetically ordered state. Validating the proposed model will require a systematic magnetic neutron diffraction study in applied magnetic fields in the future.

\section*{\label{sec:ack}Acknowledgements\protect\\}
This work is supported by a MINT grant of the Swiss National Science Foundation. We acknowledge the beam time allocation at CAMEA, PSI (ID: 20240901), as well as the support of the HLD at HZDR, member of the European Magnetic Field Laboratory (EMFL), and the W\"{u}rzburg-Dresden Cluster of Excellence on Complexity and Topology in Quantum Matter - $ct.qmat$ (EXC 2147, project No. 390858490). A portion of this research used resources at the High Flux Isotope Reactor (HFIR), a DOE Office of Science User Facility operated by the Oak Ridge National Laboratory. The beam time was allocated to the DEMAND instrument (proposal number IPTS-33733.1).

\bibliography{bibliography}% Produces the bibliography via BibTeX.

%apsrev4-2.bst 2019-01-14 (MD) hand-edited version of apsrev4-1.bst
%Control: key (0)
%Control: author (8) initials jnrlst
%Control: editor formatted (1) identically to author
%Control: production of article title (0) allowed
%Control: page (0) single
%Control: year (1) truncated
%Control: production of eprint (0) enabled
\begin{thebibliography}{27}%
\makeatletter
\providecommand \@ifxundefined [1]{%
 \@ifx{#1\undefined}
}%
\providecommand \@ifnum [1]{%
 \ifnum #1\expandafter \@firstoftwo
 \else \expandafter \@secondoftwo
 \fi
}%
\providecommand \@ifx [1]{%
 \ifx #1\expandafter \@firstoftwo
 \else \expandafter \@secondoftwo
 \fi
}%
\providecommand \natexlab [1]{#1}%
\providecommand \enquote  [1]{``#1''}%
\providecommand \bibnamefont  [1]{#1}%
\providecommand \bibfnamefont [1]{#1}%
\providecommand \citenamefont [1]{#1}%
\providecommand \href@noop [0]{\@secondoftwo}%
\providecommand \href [0]{\begingroup \@sanitize@url \@href}%
\providecommand \@href[1]{\@@startlink{#1}\@@href}%
\providecommand \@@href[1]{\endgroup#1\@@endlink}%
\providecommand \@sanitize@url [0]{\catcode `\\12\catcode `\$12\catcode
  `\&12\catcode `\#12\catcode `\^12\catcode `\_12\catcode `\%12\relax}%
\providecommand \@@startlink[1]{}%
\providecommand \@@endlink[0]{}%
\providecommand \url  [0]{\begingroup\@sanitize@url \@url }%
\providecommand \@url [1]{\endgroup\@href {#1}{\urlprefix }}%
\providecommand \urlprefix  [0]{URL }%
\providecommand \Eprint [0]{\href }%
\providecommand \doibase [0]{https://doi.org/}%
\providecommand \selectlanguage [0]{\@gobble}%
\providecommand \bibinfo  [0]{\@secondoftwo}%
\providecommand \bibfield  [0]{\@secondoftwo}%
\providecommand \translation [1]{[#1]}%
\providecommand \BibitemOpen [0]{}%
\providecommand \bibitemStop [0]{}%
\providecommand \bibitemNoStop [0]{.\EOS\space}%
\providecommand \EOS [0]{\spacefactor3000\relax}%
\providecommand \BibitemShut  [1]{\csname bibitem#1\endcsname}%
\let\auto@bib@innerbib\@empty
%</preamble>
\bibitem [{\citenamefont {Coldea}\ \emph {et~al.}(2001)\citenamefont {Coldea},
  \citenamefont {Tennant}, \citenamefont {Tsvelik},\ and\ \citenamefont
  {Tylczynski}}]{CCC1}%
  \BibitemOpen
  \bibfield  {author} {\bibinfo {author} {\bibfnamefont {R.}~\bibnamefont
  {Coldea}}, \bibinfo {author} {\bibfnamefont {D.~A.}\ \bibnamefont {Tennant}},
  \bibinfo {author} {\bibfnamefont {A.~M.}\ \bibnamefont {Tsvelik}},\ and\
  \bibinfo {author} {\bibfnamefont {Z.}~\bibnamefont {Tylczynski}},\ }\bibfield
   {title} {\bibinfo {title} {Experimental realization of a 2{D} fractional
  quantum spin liquid},\ }\href {https://doi.org/10.1103/PhysRevLett.86.1335}
  {\bibfield  {journal} {\bibinfo  {journal} {Phys. Rev. Lett.}\ }\textbf
  {\bibinfo {volume} {86}},\ \bibinfo {pages} {1335} (\bibinfo {year}
  {2001})}\BibitemShut {NoStop}%
\bibitem [{\citenamefont {Coldea}\ \emph {et~al.}(2002)\citenamefont {Coldea},
  \citenamefont {Tennant}, \citenamefont {Habicht}, \citenamefont {Smeibidl},
  \citenamefont {Wolters},\ and\ \citenamefont {Tylczynski}}]{CCC2}%
  \BibitemOpen
  \bibfield  {author} {\bibinfo {author} {\bibfnamefont {R.}~\bibnamefont
  {Coldea}}, \bibinfo {author} {\bibfnamefont {D.~A.}\ \bibnamefont {Tennant}},
  \bibinfo {author} {\bibfnamefont {K.}~\bibnamefont {Habicht}}, \bibinfo
  {author} {\bibfnamefont {P.}~\bibnamefont {Smeibidl}}, \bibinfo {author}
  {\bibfnamefont {C.}~\bibnamefont {Wolters}},\ and\ \bibinfo {author}
  {\bibfnamefont {Z.}~\bibnamefont {Tylczynski}},\ }\bibfield  {title}
  {\bibinfo {title} {Direct measurement of the spin {H}amiltonian and
  observation of condensation of magnons in the 2{D} frustrated quantum magnet
  {${\mathrm{Cs}}_{2}{\mathrm{CuCl}}_{4}$}},\ }\href
  {https://doi.org/10.1103/PhysRevLett.88.137203} {\bibfield  {journal}
  {\bibinfo  {journal} {Phys. Rev. Lett.}\ }\textbf {\bibinfo {volume} {88}},\
  \bibinfo {pages} {137203} (\bibinfo {year} {2002})}\BibitemShut {NoStop}%
\bibitem [{\citenamefont {Coldea}\ \emph {et~al.}(2003)\citenamefont {Coldea},
  \citenamefont {Tennant},\ and\ \citenamefont {Tylczynski}}]{CCC3}%
  \BibitemOpen
  \bibfield  {author} {\bibinfo {author} {\bibfnamefont {R.}~\bibnamefont
  {Coldea}}, \bibinfo {author} {\bibfnamefont {D.~A.}\ \bibnamefont
  {Tennant}},\ and\ \bibinfo {author} {\bibfnamefont {Z.}~\bibnamefont
  {Tylczynski}},\ }\bibfield  {title} {\bibinfo {title} {Extended scattering
  continua characteristic of spin fractionalization in the two-dimensional
  frustrated quantum magnet {${\mathrm{Cs}}_{2}{\mathrm{CuCl}}_{4}$} observed
  by neutron scattering},\ }\href {https://doi.org/10.1103/PhysRevB.68.134424}
  {\bibfield  {journal} {\bibinfo  {journal} {Phys. Rev. B}\ }\textbf {\bibinfo
  {volume} {68}},\ \bibinfo {pages} {134424} (\bibinfo {year}
  {2003})}\BibitemShut {NoStop}%
\bibitem [{\citenamefont {Tokiwa}\ \emph {et~al.}(2006)\citenamefont {Tokiwa},
  \citenamefont {Radu}, \citenamefont {Coldea}, \citenamefont {Wilhelm},
  \citenamefont {Tylczynski},\ and\ \citenamefont {Steglich}}]{CCC4}%
  \BibitemOpen
  \bibfield  {author} {\bibinfo {author} {\bibfnamefont {Y.}~\bibnamefont
  {Tokiwa}}, \bibinfo {author} {\bibfnamefont {T.}~\bibnamefont {Radu}},
  \bibinfo {author} {\bibfnamefont {R.}~\bibnamefont {Coldea}}, \bibinfo
  {author} {\bibfnamefont {H.}~\bibnamefont {Wilhelm}}, \bibinfo {author}
  {\bibfnamefont {Z.}~\bibnamefont {Tylczynski}},\ and\ \bibinfo {author}
  {\bibfnamefont {F.}~\bibnamefont {Steglich}},\ }\bibfield  {title} {\bibinfo
  {title} {Magnetic phase transitions in the two-dimensional frustrated quantum
  antiferromagnet {${\mathrm{Cs}}_{2}\mathrm{Cu}{\mathrm{Cl}}_{4}$}},\ }\href
  {https://doi.org/10.1103/PhysRevB.73.134414} {\bibfield  {journal} {\bibinfo
  {journal} {Phys. Rev. B}\ }\textbf {\bibinfo {volume} {73}},\ \bibinfo
  {pages} {134414} (\bibinfo {year} {2006})}\BibitemShut {NoStop}%
\bibitem [{\citenamefont {Ono}\ \emph {et~al.}(2003)\citenamefont {Ono},
  \citenamefont {Tanaka}, \citenamefont {Aruga~Katori}, \citenamefont
  {Ishikawa}, \citenamefont {Mitamura},\ and\ \citenamefont {Goto}}]{CCB1}%
  \BibitemOpen
  \bibfield  {author} {\bibinfo {author} {\bibfnamefont {T.}~\bibnamefont
  {Ono}}, \bibinfo {author} {\bibfnamefont {H.}~\bibnamefont {Tanaka}},
  \bibinfo {author} {\bibfnamefont {H.}~\bibnamefont {Aruga~Katori}}, \bibinfo
  {author} {\bibfnamefont {F.}~\bibnamefont {Ishikawa}}, \bibinfo {author}
  {\bibfnamefont {H.}~\bibnamefont {Mitamura}},\ and\ \bibinfo {author}
  {\bibfnamefont {T.}~\bibnamefont {Goto}},\ }\bibfield  {title} {\bibinfo
  {title} {Magnetization plateau in the frustrated quantum spin system
  {${\mathrm{Cs}}_{2}{\mathrm{CuBr}}_{4}$}},\ }\href
  {https://doi.org/10.1103/PhysRevB.67.104431} {\bibfield  {journal} {\bibinfo
  {journal} {Phys. Rev. B}\ }\textbf {\bibinfo {volume} {67}},\ \bibinfo
  {pages} {104431} (\bibinfo {year} {2003})}\BibitemShut {NoStop}%
\bibitem [{\citenamefont {Fortune}\ \emph {et~al.}(2009)\citenamefont
  {Fortune}, \citenamefont {Hannahs}, \citenamefont {Yoshida}, \citenamefont
  {Sherline}, \citenamefont {Ono}, \citenamefont {Tanaka},\ and\ \citenamefont
  {Takano}}]{CCB2}%
  \BibitemOpen
  \bibfield  {author} {\bibinfo {author} {\bibfnamefont {N.~A.}\ \bibnamefont
  {Fortune}}, \bibinfo {author} {\bibfnamefont {S.~T.}\ \bibnamefont
  {Hannahs}}, \bibinfo {author} {\bibfnamefont {Y.}~\bibnamefont {Yoshida}},
  \bibinfo {author} {\bibfnamefont {T.~E.}\ \bibnamefont {Sherline}}, \bibinfo
  {author} {\bibfnamefont {T.}~\bibnamefont {Ono}}, \bibinfo {author}
  {\bibfnamefont {H.}~\bibnamefont {Tanaka}},\ and\ \bibinfo {author}
  {\bibfnamefont {Y.}~\bibnamefont {Takano}},\ }\bibfield  {title} {\bibinfo
  {title} {Cascade of magnetic-field-induced quantum phase transitions in a
  spin-{$\frac{1}{2}$} triangular-lattice antiferromagnet},\ }\href
  {https://doi.org/10.1103/PhysRevLett.102.257201} {\bibfield  {journal}
  {\bibinfo  {journal} {Phys. Rev. Lett.}\ }\textbf {\bibinfo {volume} {102}},\
  \bibinfo {pages} {257201} (\bibinfo {year} {2009})}\BibitemShut {NoStop}%
\bibitem [{\citenamefont {Kenzelmann}\ \emph {et~al.}(2002)\citenamefont
  {Kenzelmann}, \citenamefont {Coldea}, \citenamefont {Tennant}, \citenamefont
  {Visser}, \citenamefont {Hofmann}, \citenamefont {Smeibidl},\ and\
  \citenamefont {Tylczynski}}]{CCoC1}%
  \BibitemOpen
  \bibfield  {author} {\bibinfo {author} {\bibfnamefont {M.}~\bibnamefont
  {Kenzelmann}}, \bibinfo {author} {\bibfnamefont {R.}~\bibnamefont {Coldea}},
  \bibinfo {author} {\bibfnamefont {D.~A.}\ \bibnamefont {Tennant}}, \bibinfo
  {author} {\bibfnamefont {D.}~\bibnamefont {Visser}}, \bibinfo {author}
  {\bibfnamefont {M.}~\bibnamefont {Hofmann}}, \bibinfo {author} {\bibfnamefont
  {P.}~\bibnamefont {Smeibidl}},\ and\ \bibinfo {author} {\bibfnamefont
  {Z.}~\bibnamefont {Tylczynski}},\ }\bibfield  {title} {\bibinfo {title}
  {Order-to-disorder transition in the {$\mathrm{XY}$}-like quantum magnet
  {${\mathrm{Cs}}_{2}{\mathrm{CoCl}}_{4}$} induced by noncommuting applied
  fields},\ }\href {https://doi.org/10.1103/PhysRevB.65.144432} {\bibfield
  {journal} {\bibinfo  {journal} {Phys. Rev. B}\ }\textbf {\bibinfo {volume}
  {65}},\ \bibinfo {pages} {144432} (\bibinfo {year} {2002})}\BibitemShut
  {NoStop}%
\bibitem [{\citenamefont {Breunig}\ \emph {et~al.}(2013)\citenamefont
  {Breunig}, \citenamefont {Garst}, \citenamefont {Sela}, \citenamefont
  {Buldmann}, \citenamefont {Becker}, \citenamefont {Bohat\'y}, \citenamefont
  {M\"uller},\ and\ \citenamefont {Lorenz}}]{CCoC2}%
  \BibitemOpen
  \bibfield  {author} {\bibinfo {author} {\bibfnamefont {O.}~\bibnamefont
  {Breunig}}, \bibinfo {author} {\bibfnamefont {M.}~\bibnamefont {Garst}},
  \bibinfo {author} {\bibfnamefont {E.}~\bibnamefont {Sela}}, \bibinfo {author}
  {\bibfnamefont {B.}~\bibnamefont {Buldmann}}, \bibinfo {author}
  {\bibfnamefont {P.}~\bibnamefont {Becker}}, \bibinfo {author} {\bibfnamefont
  {L.}~\bibnamefont {Bohat\'y}}, \bibinfo {author} {\bibfnamefont
  {R.}~\bibnamefont {M\"uller}},\ and\ \bibinfo {author} {\bibfnamefont
  {T.}~\bibnamefont {Lorenz}},\ }\bibfield  {title} {\bibinfo {title}
  {Spin-{$\frac{1}{2}$} {XXZ} chain system
  {${\mathrm{Cs}}_{2}{\mathrm{CoCl}}_{4}$} in a transverse magnetic field},\
  }\href {https://doi.org/10.1103/PhysRevLett.111.187202} {\bibfield  {journal}
  {\bibinfo  {journal} {Phys. Rev. Lett.}\ }\textbf {\bibinfo {volume} {111}},\
  \bibinfo {pages} {187202} (\bibinfo {year} {2013})}\BibitemShut {NoStop}%
\bibitem [{\citenamefont {Breunig}\ \emph {et~al.}(2015)\citenamefont
  {Breunig}, \citenamefont {Garst}, \citenamefont {Rosch}, \citenamefont
  {Sela}, \citenamefont {Buldmann}, \citenamefont {Becker}, \citenamefont
  {Bohat\'y}, \citenamefont {M\"uller},\ and\ \citenamefont {Lorenz}}]{CCoC3}%
  \BibitemOpen
  \bibfield  {author} {\bibinfo {author} {\bibfnamefont {O.}~\bibnamefont
  {Breunig}}, \bibinfo {author} {\bibfnamefont {M.}~\bibnamefont {Garst}},
  \bibinfo {author} {\bibfnamefont {A.}~\bibnamefont {Rosch}}, \bibinfo
  {author} {\bibfnamefont {E.}~\bibnamefont {Sela}}, \bibinfo {author}
  {\bibfnamefont {B.}~\bibnamefont {Buldmann}}, \bibinfo {author}
  {\bibfnamefont {P.}~\bibnamefont {Becker}}, \bibinfo {author} {\bibfnamefont
  {L.}~\bibnamefont {Bohat\'y}}, \bibinfo {author} {\bibfnamefont
  {R.}~\bibnamefont {M\"uller}},\ and\ \bibinfo {author} {\bibfnamefont
  {T.}~\bibnamefont {Lorenz}},\ }\bibfield  {title} {\bibinfo {title}
  {Low-temperature ordered phases of the spin-{$\frac{1}{2}$} {XXZ} chain
  system {${\mathrm{Cs}}_{2}{\mathrm{CoCl}}_{4}$}},\ }\href
  {https://doi.org/10.1103/PhysRevB.91.024423} {\bibfield  {journal} {\bibinfo
  {journal} {Phys. Rev. B}\ }\textbf {\bibinfo {volume} {91}},\ \bibinfo
  {pages} {024423} (\bibinfo {year} {2015})}\BibitemShut {NoStop}%
\bibitem [{\citenamefont {Povarov}\ \emph {et~al.}(2020)\citenamefont
  {Povarov}, \citenamefont {Facheris}, \citenamefont {Velja}, \citenamefont
  {Blosser}, \citenamefont {Yan}, \citenamefont {Gvasaliya},\ and\
  \citenamefont {Zheludev}}]{Povarov2020}%
  \BibitemOpen
  \bibfield  {author} {\bibinfo {author} {\bibfnamefont {K.~Y.}\ \bibnamefont
  {Povarov}}, \bibinfo {author} {\bibfnamefont {L.}~\bibnamefont {Facheris}},
  \bibinfo {author} {\bibfnamefont {S.}~\bibnamefont {Velja}}, \bibinfo
  {author} {\bibfnamefont {D.}~\bibnamefont {Blosser}}, \bibinfo {author}
  {\bibfnamefont {Z.}~\bibnamefont {Yan}}, \bibinfo {author} {\bibfnamefont
  {S.}~\bibnamefont {Gvasaliya}},\ and\ \bibinfo {author} {\bibfnamefont
  {A.}~\bibnamefont {Zheludev}},\ }\bibfield  {title} {\bibinfo {title}
  {Magnetization plateaux cascade in the frustrated quantum antiferromagnet
  {${\mathrm{Cs}}_{2}{\mathrm{CoBr}}_{4}$}},\ }\href
  {https://doi.org/10.1103/PhysRevResearch.2.043384} {\bibfield  {journal}
  {\bibinfo  {journal} {Phys. Rev. Res.}\ }\textbf {\bibinfo {volume} {2}},\
  \bibinfo {pages} {043384} (\bibinfo {year} {2020})}\BibitemShut {NoStop}%
\bibitem [{\citenamefont {Facheris}\ \emph {et~al.}(2022)\citenamefont
  {Facheris}, \citenamefont {Povarov}, \citenamefont {Nabi}, \citenamefont
  {Mazzone}, \citenamefont {Lass}, \citenamefont {Roessli}, \citenamefont
  {Ressouche}, \citenamefont {Yan}, \citenamefont {Gvasaliya},\ and\
  \citenamefont {Zheludev}}]{Facheris2022}%
  \BibitemOpen
  \bibfield  {author} {\bibinfo {author} {\bibfnamefont {L.}~\bibnamefont
  {Facheris}}, \bibinfo {author} {\bibfnamefont {K.~Y.}\ \bibnamefont
  {Povarov}}, \bibinfo {author} {\bibfnamefont {S.~D.}\ \bibnamefont {Nabi}},
  \bibinfo {author} {\bibfnamefont {D.~G.}\ \bibnamefont {Mazzone}}, \bibinfo
  {author} {\bibfnamefont {J.}~\bibnamefont {Lass}}, \bibinfo {author}
  {\bibfnamefont {B.}~\bibnamefont {Roessli}}, \bibinfo {author} {\bibfnamefont
  {E.}~\bibnamefont {Ressouche}}, \bibinfo {author} {\bibfnamefont
  {Z.}~\bibnamefont {Yan}}, \bibinfo {author} {\bibfnamefont {S.}~\bibnamefont
  {Gvasaliya}},\ and\ \bibinfo {author} {\bibfnamefont {A.}~\bibnamefont
  {Zheludev}},\ }\bibfield  {title} {\bibinfo {title} {Spin density wave versus
  fractional magnetization plateau in a triangular antiferromagnet},\ }\href
  {https://doi.org/10.1103/PhysRevLett.129.087201} {\bibfield  {journal}
  {\bibinfo  {journal} {Phys. Rev. Lett.}\ }\textbf {\bibinfo {volume} {129}},\
  \bibinfo {pages} {087201} (\bibinfo {year} {2022})}\BibitemShut {NoStop}%
\bibitem [{\citenamefont {Facheris}\ \emph {et~al.}(2023)\citenamefont
  {Facheris}, \citenamefont {Nabi}, \citenamefont {Glezer~Moshe}, \citenamefont
  {Nagel}, \citenamefont {R\~o\ om}, \citenamefont {Povarov}, \citenamefont
  {Stewart}, \citenamefont {Yan},\ and\ \citenamefont
  {Zheludev}}]{Facheris2023}%
  \BibitemOpen
  \bibfield  {author} {\bibinfo {author} {\bibfnamefont {L.}~\bibnamefont
  {Facheris}}, \bibinfo {author} {\bibfnamefont {S.~D.}\ \bibnamefont {Nabi}},
  \bibinfo {author} {\bibfnamefont {A.}~\bibnamefont {Glezer~Moshe}}, \bibinfo
  {author} {\bibfnamefont {U.}~\bibnamefont {Nagel}}, \bibinfo {author}
  {\bibfnamefont {T.}~\bibnamefont {R\~o\ om}}, \bibinfo {author}
  {\bibfnamefont {K.~Y.}\ \bibnamefont {Povarov}}, \bibinfo {author}
  {\bibfnamefont {J.~R.}\ \bibnamefont {Stewart}}, \bibinfo {author}
  {\bibfnamefont {Z.}~\bibnamefont {Yan}},\ and\ \bibinfo {author}
  {\bibfnamefont {A.}~\bibnamefont {Zheludev}},\ }\bibfield  {title} {\bibinfo
  {title} {Confinement of fractional excitations in a triangular lattice
  antiferromagnet},\ }\href {https://doi.org/10.1103/PhysRevLett.130.256702}
  {\bibfield  {journal} {\bibinfo  {journal} {Phys. Rev. Lett.}\ }\textbf
  {\bibinfo {volume} {130}},\ \bibinfo {pages} {256702} (\bibinfo {year}
  {2023})}\BibitemShut {NoStop}%
\bibitem [{\citenamefont {Facheris}\ \emph {et~al.}(2024)\citenamefont
  {Facheris}, \citenamefont {Nabi}, \citenamefont {Povarov}, \citenamefont
  {Yan}, \citenamefont {Moshe}, \citenamefont {Nagel}, \citenamefont {R\~o\
  om}, \citenamefont {Podlesnyak}, \citenamefont {Ressouche}, \citenamefont
  {Beauvois}, \citenamefont {Stewart}, \citenamefont {Manuel}, \citenamefont
  {Khalyavin}, \citenamefont {Orlandi},\ and\ \citenamefont
  {Zheludev}}]{Facheris2024}%
  \BibitemOpen
  \bibfield  {author} {\bibinfo {author} {\bibfnamefont {L.}~\bibnamefont
  {Facheris}}, \bibinfo {author} {\bibfnamefont {S.~D.}\ \bibnamefont {Nabi}},
  \bibinfo {author} {\bibfnamefont {K.~Y.}\ \bibnamefont {Povarov}}, \bibinfo
  {author} {\bibfnamefont {Z.}~\bibnamefont {Yan}}, \bibinfo {author}
  {\bibfnamefont {A.~G.}\ \bibnamefont {Moshe}}, \bibinfo {author}
  {\bibfnamefont {U.}~\bibnamefont {Nagel}}, \bibinfo {author} {\bibfnamefont
  {T.}~\bibnamefont {R\~o\ om}}, \bibinfo {author} {\bibfnamefont
  {A.}~\bibnamefont {Podlesnyak}}, \bibinfo {author} {\bibfnamefont
  {E.}~\bibnamefont {Ressouche}}, \bibinfo {author} {\bibfnamefont
  {K.}~\bibnamefont {Beauvois}}, \bibinfo {author} {\bibfnamefont {J.~R.}\
  \bibnamefont {Stewart}}, \bibinfo {author} {\bibfnamefont {P.}~\bibnamefont
  {Manuel}}, \bibinfo {author} {\bibfnamefont {D.}~\bibnamefont {Khalyavin}},
  \bibinfo {author} {\bibfnamefont {F.}~\bibnamefont {Orlandi}},\ and\ \bibinfo
  {author} {\bibfnamefont {A.}~\bibnamefont {Zheludev}},\ }\bibfield  {title}
  {\bibinfo {title} {Magnetic field induced phases and spin {H}amiltonian in
  {${\mathrm{Cs}}_{2}{\mathrm{CoBr}}_{4}$}},\ }\href
  {https://doi.org/10.1103/PhysRevB.109.104433} {\bibfield  {journal} {\bibinfo
   {journal} {Phys. Rev. B}\ }\textbf {\bibinfo {volume} {109}},\ \bibinfo
  {pages} {104433} (\bibinfo {year} {2024})}\BibitemShut {NoStop}%
\bibitem [{\citenamefont {Fischer}\ and\ \citenamefont
  {Hoppe}(1990)}]{Fischer1990}%
  \BibitemOpen
  \bibfield  {author} {\bibinfo {author} {\bibfnamefont {D.}~\bibnamefont
  {Fischer}}\ and\ \bibinfo {author} {\bibfnamefont {R.}~\bibnamefont
  {Hoppe}},\ }\bibfield  {title} {\bibinfo {title} {Zur konstitution von
  oxoruthenaten({VI}) 1. Über den aufbau von
  {${\mathrm{Cs}}_{2}{\mathrm{RuO}}_{4}$}},\ }\href
  {https://doi.org/https://doi.org/10.1002/zaac.19905910110} {\bibfield
  {journal} {\bibinfo  {journal} {Zeitschrift für anorganische und allgemeine
  Chemie}\ }\textbf {\bibinfo {volume} {591}},\ \bibinfo {pages} {87} (\bibinfo
  {year} {1990})}\BibitemShut {NoStop}%
\bibitem [{\citenamefont {Blosser}\ \emph {et~al.}(2020)\citenamefont
  {Blosser}, \citenamefont {Facheris},\ and\ \citenamefont {Zheludev}}]{fb}%
  \BibitemOpen
  \bibfield  {author} {\bibinfo {author} {\bibfnamefont {D.}~\bibnamefont
  {Blosser}}, \bibinfo {author} {\bibfnamefont {L.}~\bibnamefont {Facheris}},\
  and\ \bibinfo {author} {\bibfnamefont {A.}~\bibnamefont {Zheludev}},\
  }\bibfield  {title} {\bibinfo {title} {Miniature capacitive {F}araday force
  magnetometer for magnetization measurements at low temperatures and high
  magnetic fields},\ }\href {https://doi.org/10.1063/5.0005850} {\bibfield
  {journal} {\bibinfo  {journal} {Review of Scientific Instruments}\ }\textbf
  {\bibinfo {volume} {91}},\ \bibinfo {pages} {073905} (\bibinfo {year}
  {2020})}\BibitemShut {NoStop}%
\bibitem [{\citenamefont {Skourski}\ \emph {et~al.}(2011)\citenamefont
  {Skourski}, \citenamefont {Kuz'min}, \citenamefont {Skokov}, \citenamefont
  {Andreev},\ and\ \citenamefont {Wosnitza}}]{HMFL}%
  \BibitemOpen
  \bibfield  {author} {\bibinfo {author} {\bibfnamefont {Y.}~\bibnamefont
  {Skourski}}, \bibinfo {author} {\bibfnamefont {M.~D.}\ \bibnamefont
  {Kuz'min}}, \bibinfo {author} {\bibfnamefont {K.~P.}\ \bibnamefont {Skokov}},
  \bibinfo {author} {\bibfnamefont {A.~V.}\ \bibnamefont {Andreev}},\ and\
  \bibinfo {author} {\bibfnamefont {J.}~\bibnamefont {Wosnitza}},\ }\bibfield
  {title} {\bibinfo {title} {High-field magnetization of
  {${\mathrm{Ho}}_{2}{\mathrm{Fe}}_{17}$}},\ }\href
  {https://doi.org/10.1103/PhysRevB.83.214420} {\bibfield  {journal} {\bibinfo
  {journal} {Phys. Rev. B}\ }\textbf {\bibinfo {volume} {83}},\ \bibinfo
  {pages} {214420} (\bibinfo {year} {2011})}\BibitemShut {NoStop}%
\bibitem [{\citenamefont {Cao}\ \emph {et~al.}(2019)\citenamefont {Cao},
  \citenamefont {Chakoumakos}, \citenamefont {Andrews}, \citenamefont {Wu},
  \citenamefont {Riedel}, \citenamefont {Hodges}, \citenamefont {Zhou},
  \citenamefont {Gregory}, \citenamefont {Haberl}, \citenamefont {Molaison},\
  and\ \citenamefont {Lynn}}]{demand}%
  \BibitemOpen
  \bibfield  {author} {\bibinfo {author} {\bibfnamefont {H.}~\bibnamefont
  {Cao}}, \bibinfo {author} {\bibfnamefont {B.~C.}\ \bibnamefont
  {Chakoumakos}}, \bibinfo {author} {\bibfnamefont {K.~M.}\ \bibnamefont
  {Andrews}}, \bibinfo {author} {\bibfnamefont {Y.}~\bibnamefont {Wu}},
  \bibinfo {author} {\bibfnamefont {R.~A.}\ \bibnamefont {Riedel}}, \bibinfo
  {author} {\bibfnamefont {J.}~\bibnamefont {Hodges}}, \bibinfo {author}
  {\bibfnamefont {W.}~\bibnamefont {Zhou}}, \bibinfo {author} {\bibfnamefont
  {R.}~\bibnamefont {Gregory}}, \bibinfo {author} {\bibfnamefont
  {B.}~\bibnamefont {Haberl}}, \bibinfo {author} {\bibfnamefont
  {J.}~\bibnamefont {Molaison}},\ and\ \bibinfo {author} {\bibfnamefont
  {G.~W.}\ \bibnamefont {Lynn}},\ }\bibfield  {title} {\bibinfo {title}
  {{DEMAND}, a {D}imensional {E}xtreme {M}agnetic {N}eutron {D}iffractometer at
  the {H}igh {F}lux {I}sotope {R}eactor},\ }\href
  {https://www.mdpi.com/2073-4352/9/1/5} {\bibfield  {journal} {\bibinfo
  {journal} {Crystals}\ }\textbf {\bibinfo {volume} {9}} (\bibinfo {year}
  {2019})}\BibitemShut {NoStop}%
\bibitem [{\citenamefont {Lass}\ \emph {et~al.}(2023)\citenamefont {Lass},
  \citenamefont {Jacobsen}, \citenamefont {Krighaar}, \citenamefont {Graf},
  \citenamefont {Groitl}, \citenamefont {Herzog}, \citenamefont {Yamada},
  \citenamefont {Kägi}, \citenamefont {Müller}, \citenamefont {Bürge},
  \citenamefont {Schild}, \citenamefont {Lehmann}, \citenamefont {Bollhalder},
  \citenamefont {Keller}, \citenamefont {Bartkowiak}, \citenamefont {Filges},
  \citenamefont {Greuter}, \citenamefont {Theidel}, \citenamefont {Rønnow},
  \citenamefont {Niedermayer},\ and\ \citenamefont {Mazzone}}]{CAMEARSI}%
  \BibitemOpen
  \bibfield  {author} {\bibinfo {author} {\bibfnamefont {J.}~\bibnamefont
  {Lass}}, \bibinfo {author} {\bibfnamefont {H.}~\bibnamefont {Jacobsen}},
  \bibinfo {author} {\bibfnamefont {K.~M.~L.}\ \bibnamefont {Krighaar}},
  \bibinfo {author} {\bibfnamefont {D.}~\bibnamefont {Graf}}, \bibinfo {author}
  {\bibfnamefont {F.}~\bibnamefont {Groitl}}, \bibinfo {author} {\bibfnamefont
  {F.}~\bibnamefont {Herzog}}, \bibinfo {author} {\bibfnamefont
  {M.}~\bibnamefont {Yamada}}, \bibinfo {author} {\bibfnamefont
  {C.}~\bibnamefont {Kägi}}, \bibinfo {author} {\bibfnamefont {R.~A.}\
  \bibnamefont {Müller}}, \bibinfo {author} {\bibfnamefont {R.}~\bibnamefont
  {Bürge}}, \bibinfo {author} {\bibfnamefont {M.}~\bibnamefont {Schild}},
  \bibinfo {author} {\bibfnamefont {M.~S.}\ \bibnamefont {Lehmann}}, \bibinfo
  {author} {\bibfnamefont {A.}~\bibnamefont {Bollhalder}}, \bibinfo {author}
  {\bibfnamefont {P.}~\bibnamefont {Keller}}, \bibinfo {author} {\bibfnamefont
  {M.}~\bibnamefont {Bartkowiak}}, \bibinfo {author} {\bibfnamefont
  {U.}~\bibnamefont {Filges}}, \bibinfo {author} {\bibfnamefont
  {U.}~\bibnamefont {Greuter}}, \bibinfo {author} {\bibfnamefont
  {G.}~\bibnamefont {Theidel}}, \bibinfo {author} {\bibfnamefont {H.~M.}\
  \bibnamefont {Rønnow}}, \bibinfo {author} {\bibfnamefont {C.}~\bibnamefont
  {Niedermayer}},\ and\ \bibinfo {author} {\bibfnamefont {D.~G.}\ \bibnamefont
  {Mazzone}},\ }\bibfield  {title} {\bibinfo {title} {Commissioning of the
  novel {C}ontinuous {A}ngle {M}ulti-energy {A}nalysis spectrometer at the
  {P}aul {S}cherrer {I}nstitut},\ }\href {https://doi.org/10.1063/5.0128226}
  {\bibfield  {journal} {\bibinfo  {journal} {Review of Scientific
  Instruments}\ }\textbf {\bibinfo {volume} {94}},\ \bibinfo {pages} {023302}
  (\bibinfo {year} {2023})}\BibitemShut {NoStop}%
\bibitem [{\citenamefont {Lass}\ \emph {et~al.}(2020)\citenamefont {Lass},
  \citenamefont {Jacobsen}, \citenamefont {Mazzone},\ and\ \citenamefont
  {Lefmann}}]{MJOLNIR}%
  \BibitemOpen
  \bibfield  {author} {\bibinfo {author} {\bibfnamefont {J.}~\bibnamefont
  {Lass}}, \bibinfo {author} {\bibfnamefont {H.}~\bibnamefont {Jacobsen}},
  \bibinfo {author} {\bibfnamefont {D.~G.}\ \bibnamefont {Mazzone}},\ and\
  \bibinfo {author} {\bibfnamefont {K.}~\bibnamefont {Lefmann}},\ }\bibfield
  {title} {\bibinfo {title} {{MJOLNIR}: {A} software package for multiplexing
  neutron spectrometers},\ }\href
  {https://doi.org/https://doi.org/10.1016/j.softx.2020.100600} {\bibfield
  {journal} {\bibinfo  {journal} {SoftwareX}\ }\textbf {\bibinfo {volume}
  {12}},\ \bibinfo {pages} {100600} (\bibinfo {year} {2020})}\BibitemShut
  {NoStop}%
\bibitem [{\citenamefont {Flavi\'an}\ \emph {et~al.}(2020)\citenamefont
  {Flavi\'an}, \citenamefont {Hayashida}, \citenamefont {Huberich},
  \citenamefont {Blosser}, \citenamefont {Povarov}, \citenamefont {Yan},
  \citenamefont {Gvasaliya},\ and\ \citenamefont {Zheludev}}]{flavian}%
  \BibitemOpen
  \bibfield  {author} {\bibinfo {author} {\bibfnamefont {D.}~\bibnamefont
  {Flavi\'an}}, \bibinfo {author} {\bibfnamefont {S.}~\bibnamefont
  {Hayashida}}, \bibinfo {author} {\bibfnamefont {L.}~\bibnamefont {Huberich}},
  \bibinfo {author} {\bibfnamefont {D.}~\bibnamefont {Blosser}}, \bibinfo
  {author} {\bibfnamefont {K.~Y.}\ \bibnamefont {Povarov}}, \bibinfo {author}
  {\bibfnamefont {Z.}~\bibnamefont {Yan}}, \bibinfo {author} {\bibfnamefont
  {S.}~\bibnamefont {Gvasaliya}},\ and\ \bibinfo {author} {\bibfnamefont
  {A.}~\bibnamefont {Zheludev}},\ }\bibfield  {title} {\bibinfo {title}
  {Magnetic phase diagram of the linear quantum ferro-antiferromagnet
  {${\mathrm{Cs}}_{2}{\mathrm{Cu}}_{2}{\mathrm{Mo}}_{3}{\mathrm{O}}_{12}$}},\
  }\href {https://doi.org/10.1103/PhysRevB.101.224408} {\bibfield  {journal}
  {\bibinfo  {journal} {Phys. Rev. B}\ }\textbf {\bibinfo {volume} {101}},\
  \bibinfo {pages} {224408} (\bibinfo {year} {2020})}\BibitemShut {NoStop}%
\bibitem [{\citenamefont {Cordfunke}\ \emph {et~al.}(2010)\citenamefont
  {Cordfunke}, \citenamefont {Laan},\ and\ \citenamefont
  {Westrum~Jr.}}]{structrans}%
  \BibitemOpen
  \bibfield  {author} {\bibinfo {author} {\bibfnamefont {E.~H.~P.}\
  \bibnamefont {Cordfunke}}, \bibinfo {author} {\bibfnamefont {R.~R.}\
  \bibnamefont {Laan}},\ and\ \bibinfo {author} {\bibfnamefont {E.~F.}\
  \bibnamefont {Westrum~Jr.}},\ }\bibfield  {title} {\bibinfo {title} {The
  thermochemical and thermophysical properties of
  {${\mathrm{Cs}}_{2}{\mathrm{RuO}}_{4}$} and
  {${\mathrm{Cs}}_{2}{\mathrm{MnO}}_{4}$} at temperatures from 5 {K} to 1000
  {K}},\ }\href {https://doi.org/10.1002/chin.199245016} {\bibfield  {journal}
  {\bibinfo  {journal} {ChemInform}\ }\textbf {\bibinfo {volume} {23}}
  (\bibinfo {year} {2010})}\BibitemShut {NoStop}%
\bibitem [{\citenamefont {Wilson}\ and\ \citenamefont {Geist}(1993)}]{formfac}%
  \BibitemOpen
  \bibfield  {author} {\bibinfo {author} {\bibfnamefont {A.~J.~C.}\
  \bibnamefont {Wilson}}\ and\ \bibinfo {author} {\bibfnamefont
  {V.}~\bibnamefont {Geist}},\ }\bibfield  {title} {\bibinfo {title}
  {{I}nternational {T}ables for {C}rystallography. {V}olume {C}:
  {M}athematical, {P}hysical and {C}hemical {T}ables.},\ }\href
  {https://doi.org/https://doi.org/10.1002/crat.2170280117} {\bibfield
  {journal} {\bibinfo  {journal} {Crystal Research and Technology}\ }\textbf
  {\bibinfo {volume} {28}},\ \bibinfo {pages} {110} (\bibinfo {year}
  {1993})}\BibitemShut {NoStop}%
\bibitem [{\citenamefont {Prasad}\ \emph {et~al.}(2020)\citenamefont {Prasad},
  \citenamefont {Sadhukhan}, \citenamefont {Hansen}, \citenamefont {Felser},
  \citenamefont {Kanungo},\ and\ \citenamefont {Jansen}}]{formfac2}%
  \BibitemOpen
  \bibfield  {author} {\bibinfo {author} {\bibfnamefont {B.~E.}\ \bibnamefont
  {Prasad}}, \bibinfo {author} {\bibfnamefont {S.}~\bibnamefont {Sadhukhan}},
  \bibinfo {author} {\bibfnamefont {T.~C.}\ \bibnamefont {Hansen}}, \bibinfo
  {author} {\bibfnamefont {C.}~\bibnamefont {Felser}}, \bibinfo {author}
  {\bibfnamefont {S.}~\bibnamefont {Kanungo}},\ and\ \bibinfo {author}
  {\bibfnamefont {M.}~\bibnamefont {Jansen}},\ }\bibfield  {title} {\bibinfo
  {title} {Synthesis, crystal and magnetic structure of the spin-chain compound
  {$\mathrm{A}{\mathrm{g}}_{2}\mathrm{Ru}{\mathrm{O}}_{4}$}},\ }\href
  {https://doi.org/10.1103/PhysRevMaterials.4.024418} {\bibfield  {journal}
  {\bibinfo  {journal} {Phys. Rev. Mater.}\ }\textbf {\bibinfo {volume} {4}},\
  \bibinfo {pages} {024418} (\bibinfo {year} {2020})}\BibitemShut {NoStop}%
\bibitem [{\citenamefont {Mogare}\ \emph {et~al.}(2006)\citenamefont {Mogare},
  \citenamefont {Sheptyakov}, \citenamefont {Bircher}, \citenamefont
  {G{\"u}del},\ and\ \citenamefont {Jansen}}]{NRO}%
  \BibitemOpen
  \bibfield  {author} {\bibinfo {author} {\bibfnamefont {K.~M.}\ \bibnamefont
  {Mogare}}, \bibinfo {author} {\bibfnamefont {D.}~\bibnamefont {Sheptyakov}},
  \bibinfo {author} {\bibfnamefont {R.}~\bibnamefont {Bircher}}, \bibinfo
  {author} {\bibfnamefont {H.~U.}\ \bibnamefont {G{\"u}del}},\ and\ \bibinfo
  {author} {\bibfnamefont {M.}~\bibnamefont {Jansen}},\ }\bibfield  {title}
  {\bibinfo {title} {Neutron diffraction study of the magnetic structure of
  {${\mathrm{Na}}_{2}{\mathrm{RuO}}_{4}$}},\ }\href
  {https://api.semanticscholar.org/CorpusID:119412415} {\bibfield  {journal}
  {\bibinfo  {journal} {The European Physical Journal B - Condensed Matter and
  Complex Systems}\ }\textbf {\bibinfo {volume} {52}},\ \bibinfo {pages} {371}
  (\bibinfo {year} {2006})}\BibitemShut {NoStop}%
\bibitem [{\citenamefont {Dahlbom}\ \emph {et~al.}(2025)\citenamefont
  {Dahlbom}, \citenamefont {Zhang}, \citenamefont {Miles}, \citenamefont
  {Quinn}, \citenamefont {Niraula}, \citenamefont {Thipe}, \citenamefont
  {Wilson}, \citenamefont {Matin}, \citenamefont {Mankad}, \citenamefont
  {Hahn}, \citenamefont {Pajerowski}, \citenamefont {Johnston}, \citenamefont
  {Wang}, \citenamefont {Lane}, \citenamefont {Li}, \citenamefont {Bai},
  \citenamefont {Mourigal}, \citenamefont {Batista},\ and\ \citenamefont
  {Barros}}]{sunny}%
  \BibitemOpen
  \bibfield  {author} {\bibinfo {author} {\bibfnamefont {D.}~\bibnamefont
  {Dahlbom}}, \bibinfo {author} {\bibfnamefont {H.}~\bibnamefont {Zhang}},
  \bibinfo {author} {\bibfnamefont {C.}~\bibnamefont {Miles}}, \bibinfo
  {author} {\bibfnamefont {S.}~\bibnamefont {Quinn}}, \bibinfo {author}
  {\bibfnamefont {A.}~\bibnamefont {Niraula}}, \bibinfo {author} {\bibfnamefont
  {B.}~\bibnamefont {Thipe}}, \bibinfo {author} {\bibfnamefont
  {M.}~\bibnamefont {Wilson}}, \bibinfo {author} {\bibfnamefont
  {S.}~\bibnamefont {Matin}}, \bibinfo {author} {\bibfnamefont
  {H.}~\bibnamefont {Mankad}}, \bibinfo {author} {\bibfnamefont
  {S.}~\bibnamefont {Hahn}}, \bibinfo {author} {\bibfnamefont {D.}~\bibnamefont
  {Pajerowski}}, \bibinfo {author} {\bibfnamefont {S.}~\bibnamefont
  {Johnston}}, \bibinfo {author} {\bibfnamefont {Z.}~\bibnamefont {Wang}},
  \bibinfo {author} {\bibfnamefont {H.}~\bibnamefont {Lane}}, \bibinfo {author}
  {\bibfnamefont {Y.~W.}\ \bibnamefont {Li}}, \bibinfo {author} {\bibfnamefont
  {X.}~\bibnamefont {Bai}}, \bibinfo {author} {\bibfnamefont {M.}~\bibnamefont
  {Mourigal}}, \bibinfo {author} {\bibfnamefont {C.~D.}\ \bibnamefont
  {Batista}},\ and\ \bibinfo {author} {\bibfnamefont {K.}~\bibnamefont
  {Barros}},\ }\href {https://arxiv.org/abs/2501.13095} {\bibinfo {title}
  {Sunny.jl: A {J}ulia package for spin dynamics}} (\bibinfo {year} {2025}),\
  \Eprint {https://arxiv.org/abs/2501.13095} {arXiv:2501.13095 [quant-ph]}
  \BibitemShut {NoStop}%
\bibitem [{\citenamefont {Blazey}\ and\ \citenamefont
  {Rohrer}(1968)}]{PhysRev.173.574}%
  \BibitemOpen
  \bibfield  {author} {\bibinfo {author} {\bibfnamefont {K.~W.}\ \bibnamefont
  {Blazey}}\ and\ \bibinfo {author} {\bibfnamefont {H.}~\bibnamefont
  {Rohrer}},\ }\bibfield  {title} {\bibinfo {title} {Antiferromagnetism and the
  {M}agnetic {P}hase {D}iagram of {GdAl${\mathrm{O}}_{3}$}},\ }\href
  {https://doi.org/10.1103/PhysRev.173.574} {\bibfield  {journal} {\bibinfo
  {journal} {Phys. Rev.}\ }\textbf {\bibinfo {volume} {173}},\ \bibinfo {pages}
  {574} (\bibinfo {year} {1968})}\BibitemShut {NoStop}%
\bibitem [{\citenamefont {Liang}\ \emph {et~al.}(2015)\citenamefont {Liang},
  \citenamefont {Koohpayeh}, \citenamefont {Krizan}, \citenamefont {McQueen},
  \citenamefont {Cava},\ and\ \citenamefont {Ong}}]{CoNbO}%
  \BibitemOpen
  \bibfield  {author} {\bibinfo {author} {\bibfnamefont {T.}~\bibnamefont
  {Liang}}, \bibinfo {author} {\bibfnamefont {S.~M.}\ \bibnamefont
  {Koohpayeh}}, \bibinfo {author} {\bibfnamefont {J.~W.}\ \bibnamefont
  {Krizan}}, \bibinfo {author} {\bibfnamefont {T.~M.}\ \bibnamefont {McQueen}},
  \bibinfo {author} {\bibfnamefont {R.~J.}\ \bibnamefont {Cava}},\ and\
  \bibinfo {author} {\bibfnamefont {N.~P.}\ \bibnamefont {Ong}},\ }\bibfield
  {title} {\bibinfo {title} {Heat capacity peak at the quantum critical point
  of the transverse ising magnet {CoNb$_2$O$_6$}},\ }\href
  {https://link.aps.org/doi/10.1103/PhysRevB.101.224408} {\bibfield  {journal}
  {\bibinfo  {journal} {Nature Communications}\ }\textbf {\bibinfo {volume}
  {6}},\ \bibinfo {pages} {7611} (\bibinfo {year} {2015})}\BibitemShut
  {NoStop}%
\end{thebibliography}%

\end{document}